\documentclass[10pt]{amsart}

\usepackage{epsfig,a4,graphics,amsmath,amsthm,amssymb,amssymb,epstopdf,fullpage}

\newcommand{\beq}{\begin{equation}}  
\newcommand{\eeq}{\end{equation}}    
\newcommand{\beqa}{\begin{eqnarray}} 
\newcommand{\eeqa}{\end{eqnarray}}   
\newcommand{\bca}{\begin{cases}}
\newcommand{\eca}{\end{cases}}

\newcommand{\Dt}{\Delta t}
\newcommand{\bfu}{\mathbf{u}}
\newcommand{\bfB}{\mathbf{B}}
\renewcommand{\(}{\left(}
\renewcommand{\)}{\right)}

\newcommand{\half}{\frac{1}{2}}

\newcommand{\numF}{\mathcal{F}}

\newcommand{\s}{{\rm s}}
\newcommand{\Ma}{{\mathcal M}}
\newcommand{\MA}{{\mathcal M}_{\rm A}}
\newcommand{\cs}{c_{\rm s}}
\newcommand{\cm}{{\rm cm}}
\newcommand{\km}{{\rm km}}
\newcommand{\pc}{{\rm pc}}
\newcommand{\g}{{\rm g}}
\newcommand{\K}{{\rm K}}
\newcommand{\G}{{\rm Gauss}}
\newcommand{\pmag}{{p_{\rm m}}}
\newcommand{\pmagzero}{{p_{{\rm m},\,0}}}
\newcommand{\vect}[1]{{\mathbf{#1}}}

\def\jnl@style#1{{\rmfamily#1}}%
\def\jref@jnl#1{{\jnl@style#1}}%

\newcommand\aj{\jref@jnl{AJ}}%
\newcommand\araa{\jref@jnl{ARA\&A}}%
\newcommand\apj{\jref@jnl{ApJ}}%
\newcommand\apjl{\jref@jnl{ApJ}}%
\newcommand\apjs{\jref@jnl{ApJS}}%
\newcommand\ao{\jref@jnl{Appl.~Opt.}}%
\newcommand\apss{\jref@jnl{Ap\&SS}}%
\newcommand\aap{\jref@jnl{A\&A}}%
\newcommand\aapr{\jref@jnl{A\&A~Rev.}}%
\newcommand\aaps{\jref@jnl{A\&AS}}%
\newcommand\azh{\jref@jnl{AZh}}%
\newcommand\baas{\jref@jnl{BAAS}}%
\newcommand\jrasc{\jref@jnl{JRASC}}%
\newcommand\memras{\jref@jnl{MmRAS}}%
\newcommand\mnras{\jref@jnl{MNRAS}}%
\newcommand\pra{\jref@jnl{Phys.~Rev.~A}}%
\newcommand\prb{\jref@jnl{Phys.~Rev.~B}}%
\newcommand\prc{\jref@jnl{Phys.~Rev.~C}}%
\newcommand\prd{\jref@jnl{Phys.~Rev.~D}}%
\newcommand\pre{\jref@jnl{Phys.~Rev.~E}}%
\newcommand\prl{\jref@jnl{Phys.~Rev.~Lett.}}%
\newcommand\pasp{\jref@jnl{PASP}}%
\newcommand\pasj{\jref@jnl{PASJ}}%
\newcommand\qjras{\jref@jnl{QJRAS}}%
\newcommand\skytel{\jref@jnl{S\&T}}%
\newcommand\solphys{\jref@jnl{Sol.~Phys.}}%
\newcommand\sovast{\jref@jnl{Soviet~Ast.}}%
\newcommand\ssr{\jref@jnl{Space~Sci.~Rev.}}%
\newcommand\zap{\jref@jnl{ZAp}}%
\newcommand\nat{\jref@jnl{Nature}}%
\newcommand\iaucirc{\jref@jnl{IAU~Circ.}}%
\newcommand\aplett{\jref@jnl{Astrophys.~Lett.}}%
\newcommand\apspr{\jref@jnl{Astrophys.~Space~Phys.~Res.}}%
\newcommand\bain{\jref@jnl{Bull.~Astron.~Inst.~Netherlands}}%
\newcommand\fcp{\jref@jnl{Fund.~Cosmic~Phys.}}%
\newcommand\gca{\jref@jnl{Geochim.~Cosmochim.~Acta}}%
\newcommand\grl{\jref@jnl{Geophys.~Res.~Lett.}}%
\newcommand\jcp{\jref@jnl{J.~Chem.~Phys.}}%
\newcommand\jgr{\jref@jnl{J.~Geophys.~Res.}}%
\newcommand\jqsrt{\jref@jnl{J.~Quant.~Spec.~Radiat.~Transf.}}%
\newcommand\memsai{\jref@jnl{Mem.~Soc.~Astron.~Italiana}}%
\newcommand\nphysa{\jref@jnl{Nucl.~Phys.~A}}%
\newcommand\physrep{\jref@jnl{Phys.~Rep.}}%
\newcommand\physscr{\jref@jnl{Phys.~Scr}}%
\newcommand\planss{\jref@jnl{Planet.~Space~Sci.}}%
\newcommand\procspie{\jref@jnl{Proc.~SPIE}}%

\numberwithin{figure}{section}
\numberwithin{equation}{section}
\numberwithin{theorem}{section}

\begin{document}

\title[A robust scheme for compressible MHD]{A robust numerical scheme for highly compressible magnetohydrodynamics: Nonlinear stability, implementation and tests}

\author[K.~Waagan]{K.~Waagan} \address[Knut Waagan]{
  \newline Center for Scientific Computation and Mathematical Modeling (CSCAMM)
  \newline University of Maryland,
  \newline CSIC Building 406, 
  \newline College Park, MD 20742-3289, USA
  }
\email[]{kwaagan@cscamm.umd.edu} 
\thanks{K.~Waagan would like to thank Dr.~Michael Kn\"olker at the High Altitude Observatory in Colorado for advice and support with this work. K.~Waagan is partially supported by NSF Grants DMS07-07949, DMS10-08397, DMS10-08058, and ONR Grant N000140910385.}

\author[C.~Federrath]{C.~Federrath} \address[Christoph Federrath]{
\newline Zentrum f\"ur Astronomie der Universit\"at Heidelberg,
\newline Institut f\"ur Theoretische Astrophysik,
\newline Albert-Ueberle-Str. 2,
\newline D-69120 Heidelberg, Germany
}
\email[]{chfeder@ita.uni-heidelberg.de}
\address[Christoph Federrath, current address]{
\newline Ecole Normale Sup\'erieure de Lyon,
\newline {CRAL}, 69364 Lyon, France
}

\thanks{C.~Federrath is grateful for support from the Landesstiftung
Baden-W\"{u}rttemberg via their program International Collaboration II
under Grant P-LS-SPII/18, and has received funding from the European Research Council under the European Community's Seventh Framework Programme (FP7/2007-2013 Grant Agreement No.~247060). C.~Federrath acknowledges computational
resources from the HLRB II project Grant pr32lo at the Leibniz
Rechenzentrum Garching for running the turbulence simulations. }
\thanks{The FLASH code was developed in part by the DOE-supported Alliances Center for
Astrophysical Thermonuclear FLASHes (ASC) at the University of Chicago.}

\author[C.~Klingenberg]{C.~Klingenberg} \address[Christian Klingenberg]{
\newline Department of Mathematics,
\newline W\"urzburg University,
\newline Am Hubland 97074 W\"urzburg, Germany}
\email[]{klingenberg@mathematik.uni-wuerzburg.de}

\date{\today}    

\begin{abstract}
The ideal MHD equations are a central model in astrophysics, and their solution relies upon stable numerical schemes. We present an implementation of a new method, which possesses excellent stability properties. Numerical tests demonstrate that the theoretical stability properties are valid in practice with negligible compromises to accuracy. The result is a highly robust scheme with state-of-the-art efficiency.  The scheme's robustness is due to entropy stability, positivity and properly discretised Powell terms. The implementation takes the form of a modification of the MHD module in the FLASH code, an adaptive mesh refinement code. We compare the new scheme with the standard FLASH implementation for MHD. Results show comparable accuracy to standard FLASH with the Roe solver, but highly improved efficiency and stability, particularly for high Mach number flows and low plasma $\beta$. The tests include 1D shock tubes, 2D instabilities and highly supersonic, 3D turbulence. We consider turbulent flows with RMS sonic Mach numbers up to 10, typical of gas flows in the interstellar medium. We investigate both strong initial magnetic fields and magnetic field amplification by the turbulent dynamo from extremely high plasma $\beta$. The energy spectra show a reasonable decrease in dissipation with grid refinement, and at a resolution of $512^3$ grid cells we identify a narrow inertial range with the expected power-law scaling. The turbulent dynamo exhibits exponential growth of magnetic pressure, with the growth rate twice as high from solenoidal forcing than from compressive forcing. Two versions of the new scheme are presented, using relaxation-based 3-wave and 5-wave approximate Riemann solvers, respectively. The 5-wave solver is more accurate in some cases, and its computational cost is close to the 3-wave solver.
\end{abstract}

\maketitle

\section{Introduction}
In order to model complex nonlinear astrophysical phenomena, numerical solutions of the equations of ideal MHD are central. Examples include the dynamics of stellar atmospheres, star formation and accretion discs. Magnetised gas flows in astrophysics are typically highly compressible, nonlinear and often supersonic. Hence, obtaining stable numerical results in such regimes is extremely challenging if the numerical
scheme must also be both accurate and efficient. Here we present a new MHD solver that preserves stable, physical solutions to
the compressible MHD equations by construction, while at the same time
showing improved efficiency and comparable accuracy to standard schemes based on the very accurate, but less stable Roe approximate Riemann solver \cite{Roe1981}.

We consider numerical solutions of the ideal MHD system (here written in conservation form, and in three dimensions, letting $I_3$ denote the $3\times3$ identity matrix)
\begin{equation}\begin{array}{l}
	 \rho_t + \nabla\cdot(\rho \bfu)=0,\\
 (\rho \bfu)_t + \nabla\cdot(\rho \bfu\otimes \bfu + (p+ \frac{1}{2} |\bfB|^2)\, {I_3} - \bfB\otimes \bfB)=0,\\
 E_t + \nabla\cdot[(E+p+\frac{1}{2} |\bfB|^2)\bfu - (\bfB\cdot \bfu)\bfB]=0,\\
 \bfB_t + \nabla\cdot(\bfB\otimes \bfu - \bfu\otimes \bfB)=0,\\
 \nabla\cdot\bfB=0,
	\label{eq:MHDP}
	\end{array}
\end{equation}
with the specific internal energy $e$, such that the total energy density is given by $E=\rho e + \half\rho\bfu^2 + \half|\bfB|^2$, and the pressure given by the equation of state $p=p(\rho,e)$. The Cartesian components of the velocity are denoted by $\bfu=(u,v,w)$. The system fits the generic form of a conservation law $U_t + \nabla\cdot \mathbf{F}(U)=0$, except for the restriction on $\nabla\cdot\bfB$. However, if this restriction is satisfied at the initial time $t=0$, it automatically holds at later times  $t>0$ for the exact solution. Shock conditions are common in astrophysics, hence \eqref{eq:MHDP} should be understood in a weak sense. The lack of regularity and large scale ranges in astrophysical flows make it particularly challenging to devise numerically stable schemes. Recent developments in nonlinear stability analysis have made such schemes possible, however. The central stability notions are entropy stability and positivity of mass density $\rho$ and internal energy $\rho e$. This paper presents an implementation of the robust positive second order scheme from \cite{W1}, which uses the entropy stable approximate Riemann solvers of \cite{BKW1, BKW2}.

Several codes are available for high-performance compressible MHD simulations. We mention Athena \cite{Athena2008}, AstroBEAR \cite{Astrobear2009}, Chombo \cite{Chombo2005}, Enzo \cite{Enzo2008}, FLASH \cite{FLASH}, Nirvana \cite{Nirvana2005}, Pluto \cite{Pluto2007}, RAMSES \cite{RAMSES2006} and VAC \cite{VAC1996}. The positive scheme of \cite{W1} has so far been unavailable in such codes. We have implemented it as an alternative MHD module in the FLASH code.

We focus here on a comparison with the original FLASH code. Benchmark tests in one and two spatial dimensions are presented. Transition to unstable turbulence-like flow is particularly emphasised. Additional tests of the algorithms may be found in \cite{BKW2, KW2010} and \cite{W1}. As an example application we consider turbulence in molecular clouds, which is characterised by high Mach numbers and the presence of magnetic fields. Before presenting the results, we give a review of the numerical schemes and their theoretical properties.

\section{Numerical schemes}
In this section we give a quick summary of the numerical techniques used, and their justification. Consider first a system of conservation laws in one spatial dimension (we consider the $x$-dimension) $U_t + F(U)_x=0$. The numerical schemes we will use are of the finite volume type
\beq
\mathcal{S}_{\Dt} U_i = U_i-  \frac{\Dt}{h}\( \numF_{i+\half}  -  \numF_{i-\half} \),
\label{eq:Godunov}
\eeq
where $U_i$ are averages over intervals (or 'cells') of length $h$ indexed by $i$ at a given time $t$, and the operator $\mathcal{S}_{\Dt}$ updates the cell averages to time $t+\Dt$. The numerical fluxes $\numF$ are evaluated at the cell interfaces, hence $U$ is conserved and we call the scheme conservative. 

\subsection{First order accuracy in one space dimension}
\label{first_order}
First order accurate schemes can be given by
\beq
\numF_{i+\half} = \numF(U_i,U_{i+1}).
\label{eq:1order}
\eeq
Typically, $\numF(\cdot,\cdot)$ is given by an approximate Riemann solver \cite{Bouchutbook,Torobook,LeVequebook}. The following are considered here: The Roe solver and the HLLE (Harten, Lax, van Leer and Einfeldt) solver as implemented in FLASH, and the multi-wave HLL-type solvers of \cite{BKW1,BKW2} (named HLL3R and HLL5R, or HLLxR).  Other HLL-type solvers  are given in \cite{HLL,Miyoshi5wave,GurskiHLLCMHD,LiHLLCMHD,LindePhD,FMMRWPowell} among others. The numerical flux should be consistent (i.e. $\numF(U,U)=F(U))$, and satisfy appropriate stability criteria. For gas and plasma dynamical systems a strong and physically meaningful stability criterion is implied by the second law of thermodynamics, which reads in one dimension:
\beq
(\rho s)_t +  (\rho u  s)_x \geq 0,
\label{entropy}
\eeq
$s$ being the specific entropy of the system. This inequality holds, by definition, for an exact Riemann solver, and should also hold for an approximate Riemann solver. The HLLxR solvers were proved in \cite{BKW1,BKW2} to satisfy \eqref{entropy}, and consequently the resulting conservative scheme satisfies a discrete version of \eqref{entropy}. We refer to an approximate Riemann solver satisfying \eqref{entropy} as being entropy stable. For the Roe and HLLE solvers there are no rigorous proofs of entropy stability. For HLLE this appears not to be a problem, while Roe solvers yield non-physical shocks at sonic points. Therefore, Roe solvers are typically equipped with a so-called entropy-fix, which adds extra dissipation at sonic points. The Roe solver also tends to produce negative mass density values, while HLLE and HLLxR provably preserve positive mass density. Positivity of mass density and entropy stability imply that the internal energy remain positive. The positivity of density and internal energy ensures that the hyperbolicity of the equations does not break down (in which case the numerical schemes no longer make sense), and implies that the computed conserved quantities are stable in $L^1$ \cite{Perthame:MUSCL}. Entropy stability and positivity are referred to as nonlinear stability conditions since they apply directly to nonlinear systems. The 'R' in HLLxR comes from the derivation of the solvers from a relaxation approximation to the MHD equations, hence they are referred to as relaxation solvers. The underlying relaxation approximation has formally  been found to be entropy dissipative through a Chapman-Enskog analysis \cite{BKW1}. The corresponding relaxation solver for hydrodynamics \cite{BouchutBGK}, was tested in an astrophysical turbulence setting in \cite{KSW2007}.

Approximate Riemann solvers tend to vary in accuracy according to the level of detail they take into account. Typically, some wave modes are better resolved than others. For the solvers considered here, the HLLE solver can only optimally resolve the fast waves, while the Roe solver can, in principle, resolve all isolated waves as well as the exact solver. The relaxation solvers lie somewhere in between: The HLL3R can optimally resolve material contacts and tangential discontinuities at vanishing Alfv{\'e}n speed, and the HLL5R additionally resolves velocity shears at vanishing Alfv{\'e}n speed. Hence HLL5R is a true generalisation of the HLLC solver for the Euler equations. These differences in resolution most prominently appear when the wave modes in question are close to stationary, i.e. much slower than $h/\Delta t$.

\subsection{Second order accuracy in one space dimension}
In order to obtain second-order accuracy, some interpolation, i.e. reconstruction of the states from cell averages, has to be employed. We consider here the conservative MUSCL-Hancock scheme, introduced in \cite{Leer:MH} (see also \cite{Torobook}). We refer to \cite{W1} for the non-conservative case. Let $W$ denote the primitive state variables $(\rho,\bfu,\bfB,p)$. Away from discontinuities the equations can be rewritten as $W_t+A(W)W_x=0$ for a matrix $A(W)$. The MUSCL-Hancock algorithm goes as follows:
\begin{enumerate}
\item
Reconstruction: evaluate discrete differences $DW_i$.  For oscillation control, we use the MC-limiter (monotonised central limiter), so for each component of $W_i$ we take
\beq
DW_i = \sigma_i \min \(2|W_{i+1}-W_i|, \half|W_{i+1}-W_{i-1}|, 2|W_{i}-W_{i-1}| \) 
\label{eq:mclimiter}
\eeq
with
\beq
\sigma_i=\bca
1, \quad W_{i+1}-W_i>0,\, W_{i}-W_{i-1}>0\\
-1, \quad W_{i+1}-W_i<0,\, W_{i}-W_{i-1}<0\\
0, \quad\text{otherwise}.
\eca
\eeq
\item
Prediction step: evaluate
\beq
W^c_i=W_i - \frac{\Dt}{2h}A(W_i)DW_i 
\label{eq:predW}
\eeq
\item
Evaluate the cell edge values
\beq
W^-_i=W^c_i-\half DW_i,\quad W^+_i=W^c_i+\half DW_i.
\label{Wrecons}
\eeq
\item
Use the cell edge values as input to the numerical flux in the conservative scheme,
\beq
\mathcal{S}_{\Dt} U_i = U_i-  \frac{\Dt}{h}\( \numF\( U^+_i, U^-_{i+1}\)  -  \numF\( U^+_{i-1}, U^-_i\) \),
\label{eq:MH}
\eeq
where $U$ denotes the conserved state variables.
\end{enumerate}
Using primitive variables as the basis for reconstructing the states was recommended for example in \cite{ppmpaper}. It ensures that material contact discontinuities are reproduced exactly (and also shear waves at vanishing Alfv{\'e}n speed). Alternatively, one can use that the primitive form $W_t+A(W)W_x=0$ may be diagonalised as
\beq
(R_i^j)_t + \lambda^j (R_i^j)_x=0, \quad j=1,2,\ldots, d
\eeq
with $W_i=X_iR_i$. The matrix $X_i$ is given by the eigenvectors $X^j_i$ of $A(W_i)$. The reconstructed gradients may then be evaluated using the relation $DW_i=X_i DR_i$, given gradients $DR_i$, and we get
\beq
W_i^{\pm} = X_i R_i^{\pm} = X_i (R_i - \frac{\Dt}{2h}\lambda R_i \pm\half DR_i) = W_i - \frac{\Dt}{2h}A(W_i)DW_i \pm \half DW_i,
\label{charrec}
\eeq
where $\lambda$ is the diagonal matrix having the eigenvalues $\lambda^j$ as entries. The original FLASH code uses the characteristic variable-based reconstruction, while we use the primitive variables with the relaxation solvers.

The limiter \eqref{eq:mclimiter} forces the scheme to behave like the first order scheme near a discontinuity, hence ensuring that the entropy dissipation of the approximate Riemann solver kicks in. Even so, it turns out to be important to ensure the positivity of mass density and internal energy. In \cite{W1}, a limiting procedure was presented such that the MUSCL-Hancock scheme is positive for any positive approximate Riemann solver. It consists of a modification that applies to any piecewise linear reconstruction of the $W$ or $R$ -variables. This limiter (more specifically the '$W$-reconstruction' of \cite{W1}) is used with the relaxation solvers.

\subsection{Multidimensionality}
Clearly our scheme is based on the treatment of one-dimensional Riemann problems. Multidimensional Riemann problems are much too complicated to be used as a basis of schemes, so various means have been suggested to derive multidimensional schemes from one-dimensional schemes. We only consider Cartesian grids here, where one can rely on dimensional splitting. This method consists of first solving the system ignoring the $y$ and $z$- derivatives, then ignoring the $x$- and $z$-derivatives etc. The splitting adds an error, and in order to maintain second order accuracy, the order of the directions must be reversed between each time step (Strang splitting). Unsplit methods have been developed to eliminate this error completely (e.g. in \cite{LDF2009,LD2009}). We limit our discussion to the split scheme, and focus on issues specific to the multidimensional MHD equations. Note however that these issues, as well as the proposed solutions, are not specific to split schemes.

Due to considering the directions separately, we have to make sense of one-dimensional problems where the longitudinal magnetic field component is non-constant, in other words we can no longer assume $\nabla\cdot\bfB = 0$ initially. In order to handle the resulting more general one-dimensional problems, we use the following approach of \cite{Powell8wave}, consisting of modifying the evolution equation for $\bfB$ to
\begin{equation}\begin{array}{l}
 \bfB_t + \nabla\cdot(\bfB\otimes \bfu - \bfu\otimes \bfB) - \bfu \nabla\cdot\bfB=0.\\
	\label{eq:BPowell}
	\end{array}
\end{equation}
Equation \eqref{eq:BPowell} implies that
\beq
(\nabla\cdot\bfB)_t + \nabla\cdot(\bfu\nabla\cdot\bfB) = 0,
\eeq
consequently $\nabla\cdot\bfB$ remains zero analytically, and numerical errors in $\nabla\cdot\bfB$ are transported by the flow. In the original FLASH code, the momentum and energy equations were also modified, which is what was actually recommended in \cite{Powell8wave}. We found it sufficient for stability in \cite{BKW2} and \cite{W1} to only include the Powell term for the induction equation. 

The one-dimensional Powell system is only conservative when $(B_n)$ is constant, so the schemes must take the form
\beq
\mathcal{S}_{\Dt} U_i = U_i - \frac{\Dt}{h}\( \mathcal{F}_l(U_i^+, U_{i+1}^-) -  \mathcal{F}_r(U_{i-1}^+, U_i^-)\) -  \Dt S_i(U_{i-1},U_i,U_{i+1}).
\eeq
In the original FLASH code, the source contributions from the cell edges are ignored, and the Powell source term is simply evaluated by a central discretisation. In \cite{BKW2} and \cite{W1}, we found that it was essential for stability, both practically and theoretically, to discretise the Powell term in a proper upwind manner. This was carried out in \cite{BKW1} for the first order scheme by extending the relaxation system to the Powell system, see also \cite{FMMRWPowell}. The second order extension can be found in \cite{W1}, and numerical tests demonstrating the importance of including a properly discretised Powell source are in \cite{FMMRWPowell,KW2010} and \cite{W1}.

We include divergence cleaning in the form of the parabolic cleaning method of \cite{Marder:divB}; see also \cite{Dedner:divB}. We use it as already implemented in FLASH. It has the advantage of not expanding the numerical stencil of the scheme. FLASH also has the non-local projection method of \cite{Brackbill:Barnes} implemented, which could in principle be used also with the new schemes. An alternative to using \eqref{eq:BPowell} is the staggered mesh (or constrained transport) approach (reviewed in \cite{TothdivB}), which in one dimension essentially means evaluating $B_n$ at the cell interface instead of as a cell average. The numerical stability results we rely on (entropy stability and positivity) are not available for staggered methods, but staggered schemes have the advantage of guaranteeing that $\nabla\cdot\bfB=0$ to approximation order in smooth regions. Note that all approximate Riemann solvers we consider may also be used as components of staggered methods.

\subsection{Isothermal gas}
As a simple model of cooling (e.g., due to radiation), it is common to assume that the gas temperature is constant. This means that $p=c_s^2\rho$, where the constant $c_s$ is the sound speed. A practical way to implement isothermality consists of (i) solving the full equations for ideal gas with a $\gamma$ close to 1 within each time step, and (ii) projecting to the isothermal state between each time step. We use $\gamma=1.00001$. This projection method is preferred due its simplicity, accuracy, and generalisability. By generalisability we mean that the projection to $p=c_s^2\rho$ is easily replaced with more elaborate cooling models. The projection method requires robust underlying numerics, since the full energy equation has to be solved properly.

\subsection{The schemes}
The changes made to the original FLASH code in order to implement the new scheme are as described above. The time step size is chosen from the maximum wave speed of the previous time step and the maximum of $\sqrt{c_s^2 + \frac{\bfB^2}{\rho}}$ over all cell averages ($c_s$ denoting sound speed). We use a CFL-number of 0.8 throughout, unless otherwise noted. We used version 2.5 of FLASH, but it is straightforward to implement the new scheme in other finite volume codes.

For numerical testing, we compare the new schemes \textsf{HLL3R} and \textsf{HLL5R} with the basic FLASH schemes for MHD. FLASH offers both a Roe and an HLLE approximate Riemann solver. The relevant schemes are summarised in Table~\ref{tab:schemes}. All schemes are finite volume schemes of MUSCL-Hancock type, with differences in the (approximate) Riemann solver, reconstruction method and Powell terms as described above. The names of the schemes are taken from the Riemann solvers for simplicity, although other algorithmic differences may also be important. To distinguish the schemes from the Riemann solvers, we use the `sans serif' font to indicate the scheme and normal font type for the Riemann solvers (see Tab.~\ref{tab:schemes}).
\begin{table}
   \centering
   \begin{tabular}{@{} lccc @{}} 
        Scheme  & \textsf{HLL3R}/\textsf{HLL5R} & \textsf{FLASH-Roe} & \textsf{FLASH-HLLE} \\
      \hline
      Riemann solver      & HLL3R/HLL5R  & Roe & HLLE \\
      Reconstruction       & positive, primitive  &  characteristic & characteristic  \\
      Powell term       & $B$ only  & full & full \\
      Powell term discretisation   &  upwind & central & central \\
     \end{tabular}
   \caption{Summary of the schemes. }
   \label{tab:schemes}
\end{table}
The thermodynamics are either adiabatic with an ideal gas equation of state or isothermal.


\section{Numerical studies 1D}
In one spatial dimension, the differences between the schemes are given mainly by the accuracy of the approximate Riemann solver and the stability properties of both the approximate Riemann solver and the reconstruction.

\subsection{Brio-Wu shock tube}
\begin{figure}
   \centering
   \includegraphics[scale=.4]{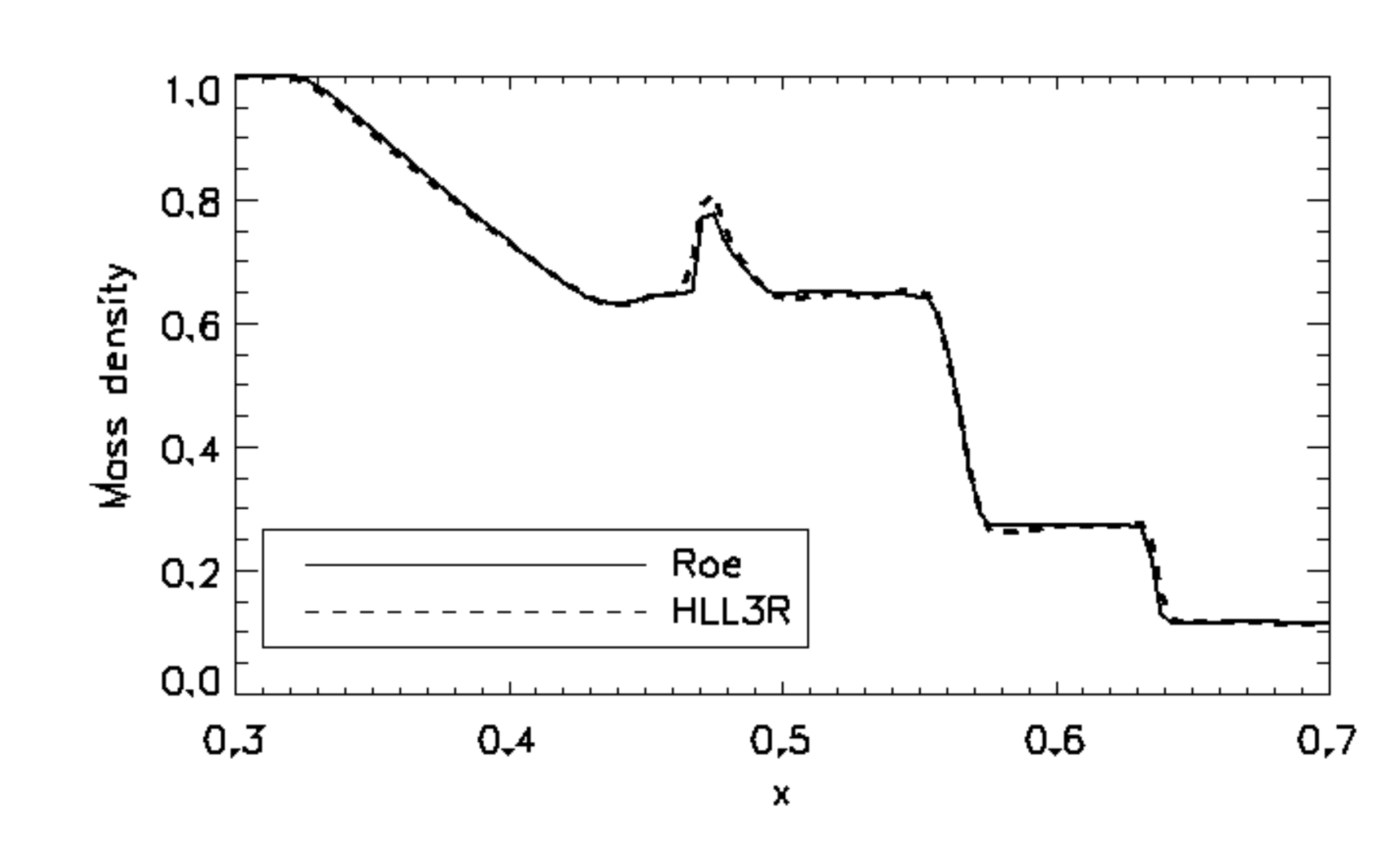} \\
   \includegraphics[scale=.4]{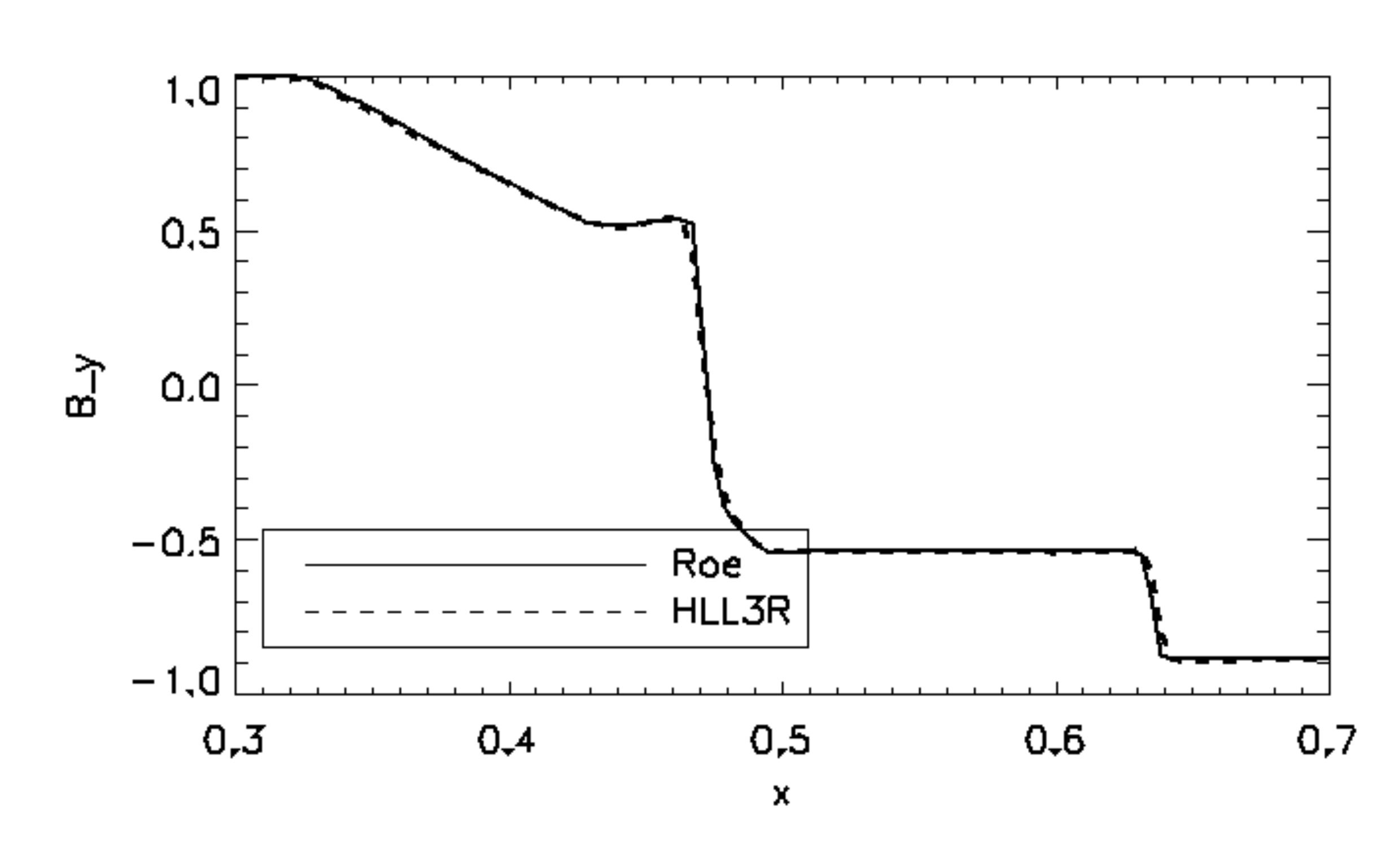} 
      \caption{Brio-Wu shock tube. Resolution $h=256^{-1}$, time $t=0.1$. \textsf{FLASH-Roe} and \textsf{HLL3R} are compared. The waves seen are from left to right, a fast rarefaction, a slow mode compound wave, a material contact discontinuity and a slow shock. The smaller amplitude right-going Alfv{\'e}n and fast waves are omitted for clarity.}
   \label{fig:bw1}
\end{figure}

\begin{figure}
   \centering
   \includegraphics[scale=.4]{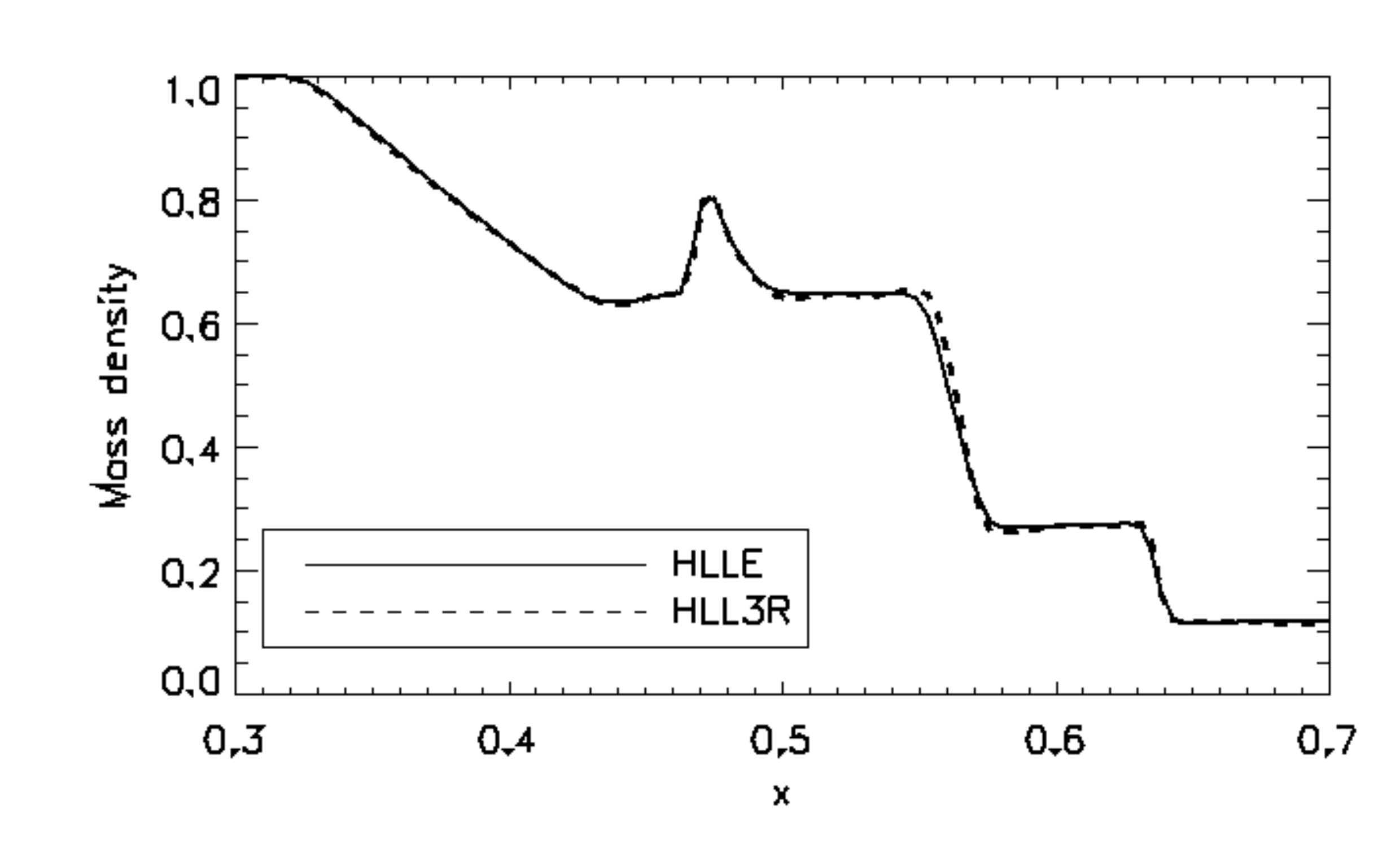} \\
   \caption{The same experiment as Figure \ref{fig:bw1}. \textsf{FLASH-HLLE} and \textsf{HLL3R} are compared.}
   \label{fig:bw1hlle}
\end{figure}

We first consider initial shock tube data from \cite{BrioWu} that have become a standard test case. It contains a variety of numerically challenging 
wave types occurring in MHD flows, in particular a fast rarefaction, a compound wave, a material contact, and a slow shock. The initial data are given by $U=U_l$ for $x<0.5$, and $U=U_r$ for $x>0.5$, with $\gamma=5/3$ and
\begin{align*}
\rho_l=1, \bfu_l=0, \bfB_l=(0.75,1,0), p_l=1\\
\rho_r=0.125, \bfu_r=0,\bfB_r=(0.75,-1,0), p_r=0.1.
\end{align*}
Figure~\ref{fig:bw1} shows small differences between \textsf{HLL3R} and \textsf{FLASH-Roe} on this standard test with \textsf{FLASH-Roe} slightly sharper on the compound wave. \textsf{FLASH-HLLE} is similar to \textsf{HLL3R}, except that the material contact was more smeared out than in \textsf{HLL3R} (see Fig.~\ref{fig:bw1hlle}).
We conclude that all the schemes perform reasonably well on this test case, with some differences in accuracy depending on the level of detail in the approximate Riemann solver.

\subsection{Low $\beta$ expansion waves}
\begin{figure}
   \centering
   \includegraphics[scale=.4]{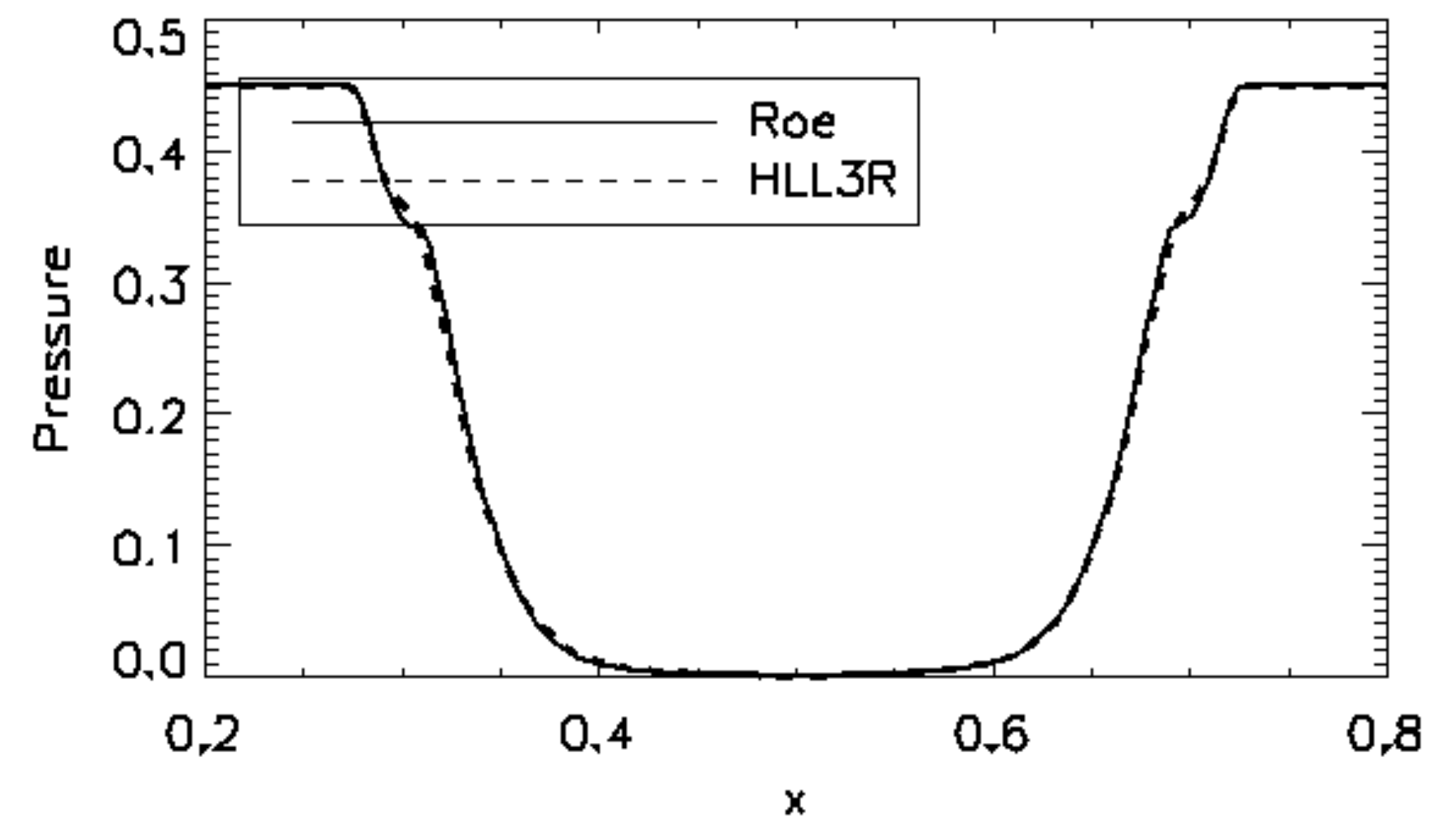}
   \caption{Low $\beta$ expansion wave. Resolution $h=256^{-1}$. Adiabatic case, velocity $\hat{u}=3.1$ at time $t=0.05$. \textsf{FLASH-Roe} and \textsf{HLL3R} are compared.}
   \label{fig:exp}
\end{figure}

This is a test from \cite{BKW2}. It consists of rarefactions into a region of low plasma $\beta$ (defined as $\beta=\frac{2p}{\bfB^2}$). The initial data are given by $U=U_l$ for $x<0.5$, and $U=U_r$ for $x>0.5$, with $\gamma=5/3$ and
\begin{align*}
\rho_l=1, \bfu_l=(-\hat{u},0,0), \bfB_l=(1,0.5,0),p_l=0.45\\
\rho_r=1, \bfu_r=(\hat{u},0,0),\bfB_r=(1,0.5,0), p_r=0.45.
\end{align*}
Hence the sound speed is initially $\sqrt{3}/2\approx 0.87$. At $\hat{u}=3.1$ there are only small differences between \textsf{FLASH-Roe} and \textsf{HLL3R} (Fig.~\ref{fig:exp}). 

\begin{figure}
   \centering
   \includegraphics[scale=.4]{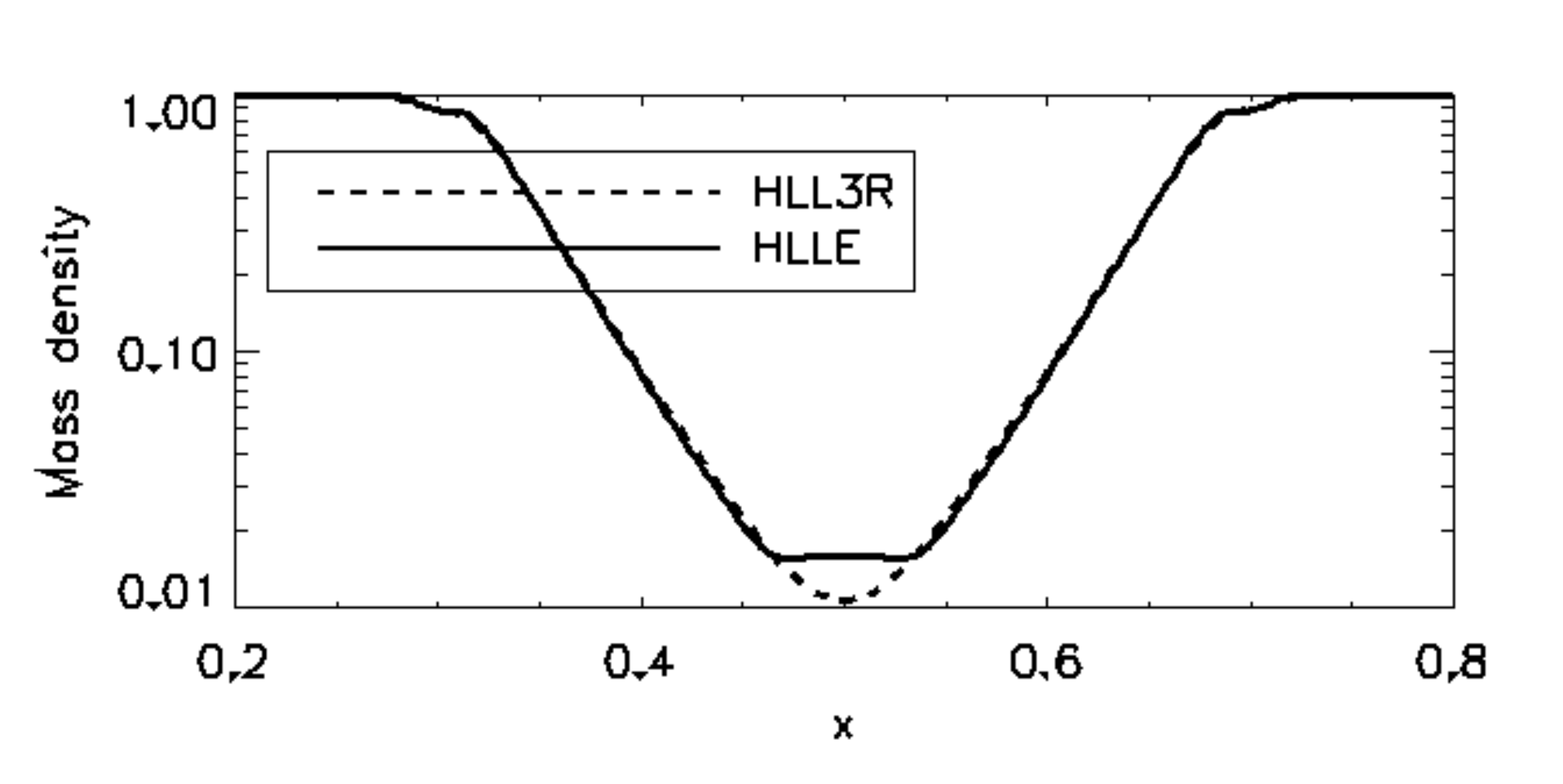} 
   \caption{Low $\beta$ expansion wave, isothermal gas. Resolution $h=256^{-1}$. Velocity $\hat{u}=3.1$ at time $t=0.05$. \textsf{FLASH-HLLE} and \textsf{HLL3R} are compared.}
   \label{fig:expisoT}
\end{figure}

We also performed this test for the isothermal case with $c_s^2=0.45$. \textsf{FLASH-Roe} could only handle $\hat{u}$-values up to about 2.9. For higher values it failed to run due to negative pressure and density. In contrast, \textsf{FLASH-HLLE} and \textsf{HLL3R} proved quite stable also in the isothermal case. They are compared in Figure~\ref{fig:expisoT}. \textsf{HLL3R} is able to represent lower densities than \textsf{FLASH-HLLE} during the expansion.

\section{Numerical studies 2D}
In two dimensions we consider two standard test cases from the numerical literature and a Kelvin-Helmholtz instability. The tests exhibit strong shock waves and complicated wave interactions. Regarding numerical stability, in addition to the approximate Riemann solver and the reconstruction, the influence of different discretisations of the Powell term are investigated.

\subsection{Orszag-Tang}
\begin{figure}
   \centering
   \includegraphics[scale=.32]{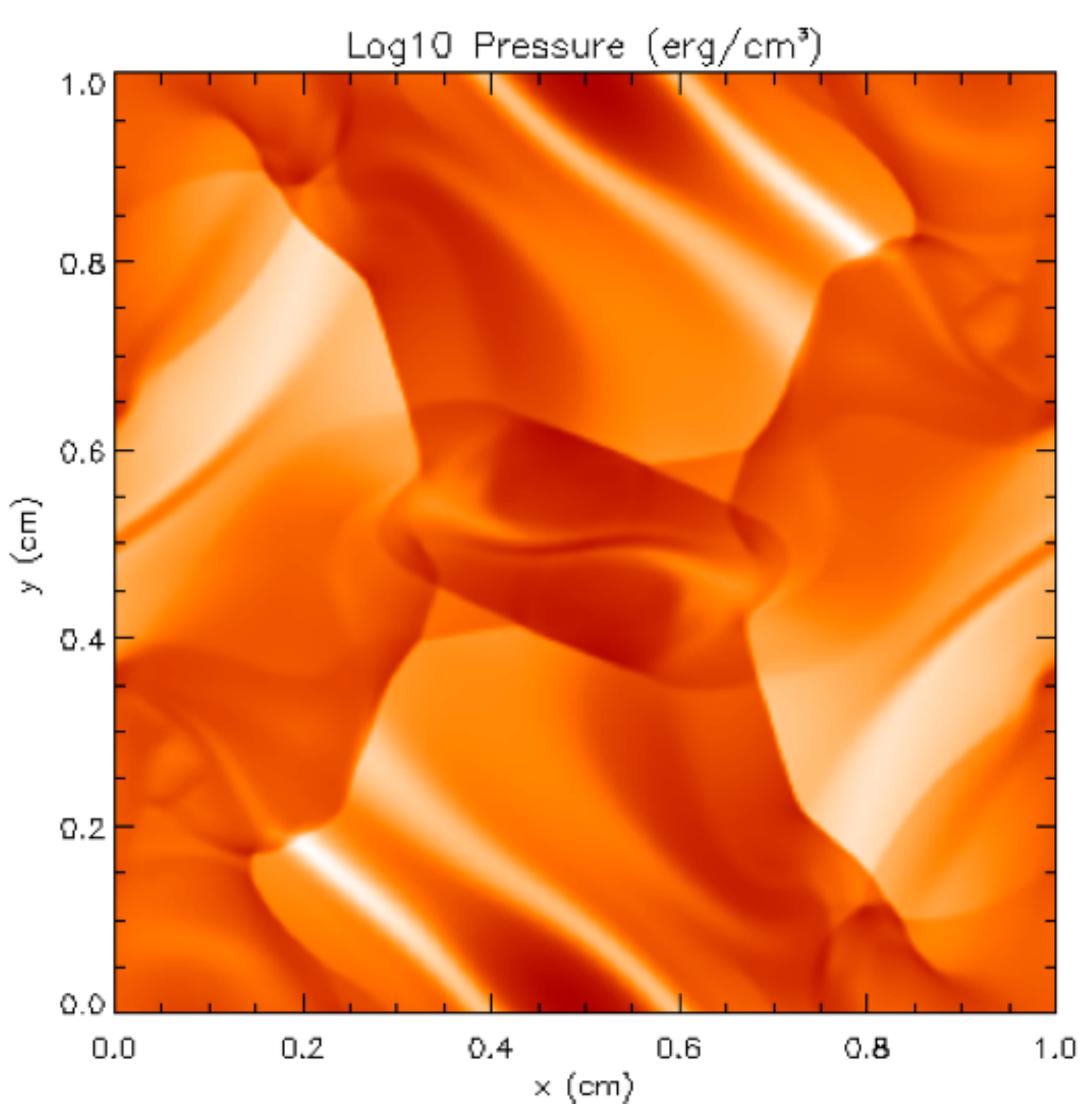} 
   \includegraphics[scale=.32]{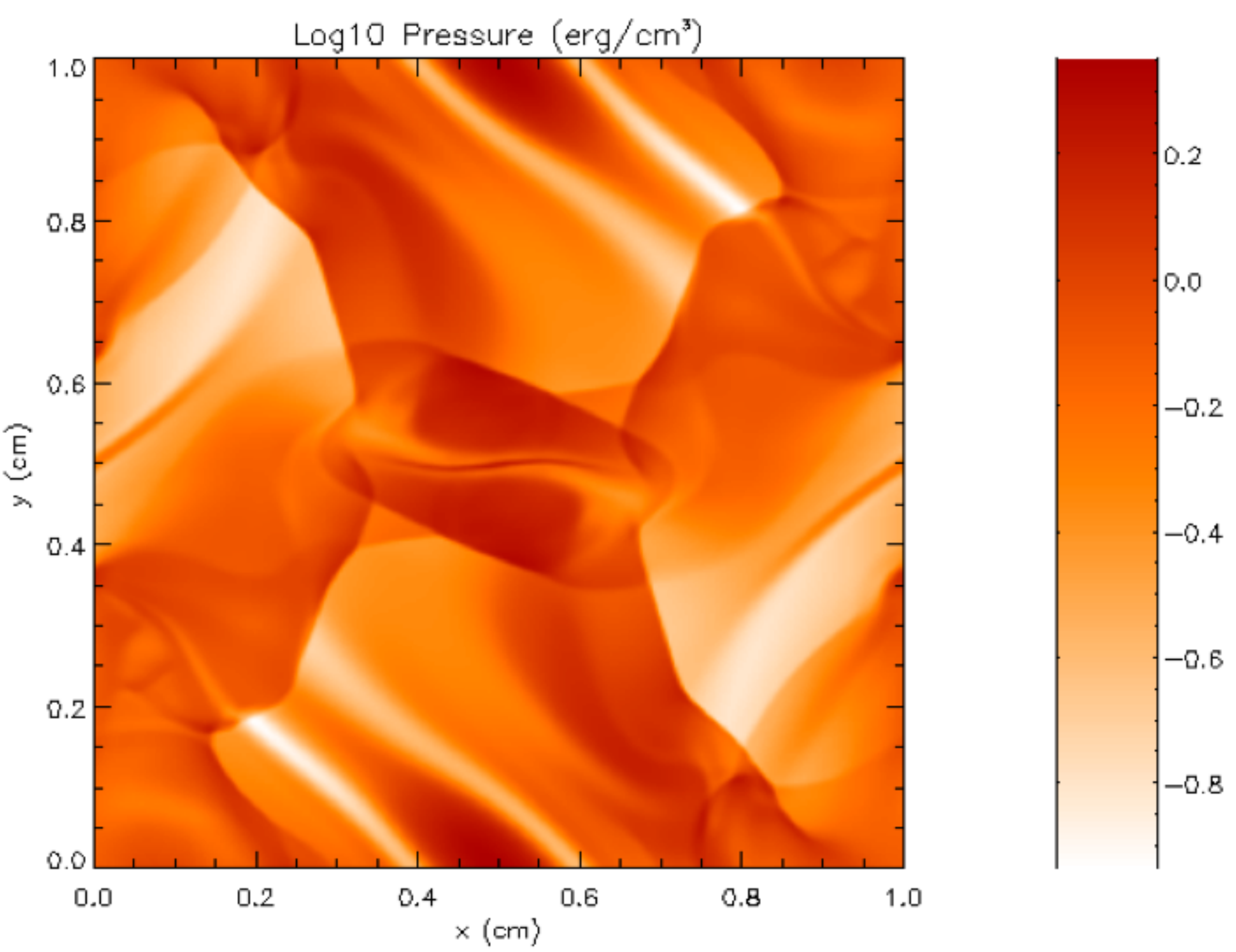} 
   \caption{Orszag-Tang test. Pressure at time $t=0.5$, with resolution $h=256^{-1}$. Left: \textsf{FLASH-HLLE}, right: \textsf{HLL3R}. The result from \textsf{FLASH-Roe} was visually hard to discern from \textsf{HLL3R}.}
   \label{fig:ot}
\end{figure}

\begin{figure}
   \centering
   \includegraphics[scale=.4]{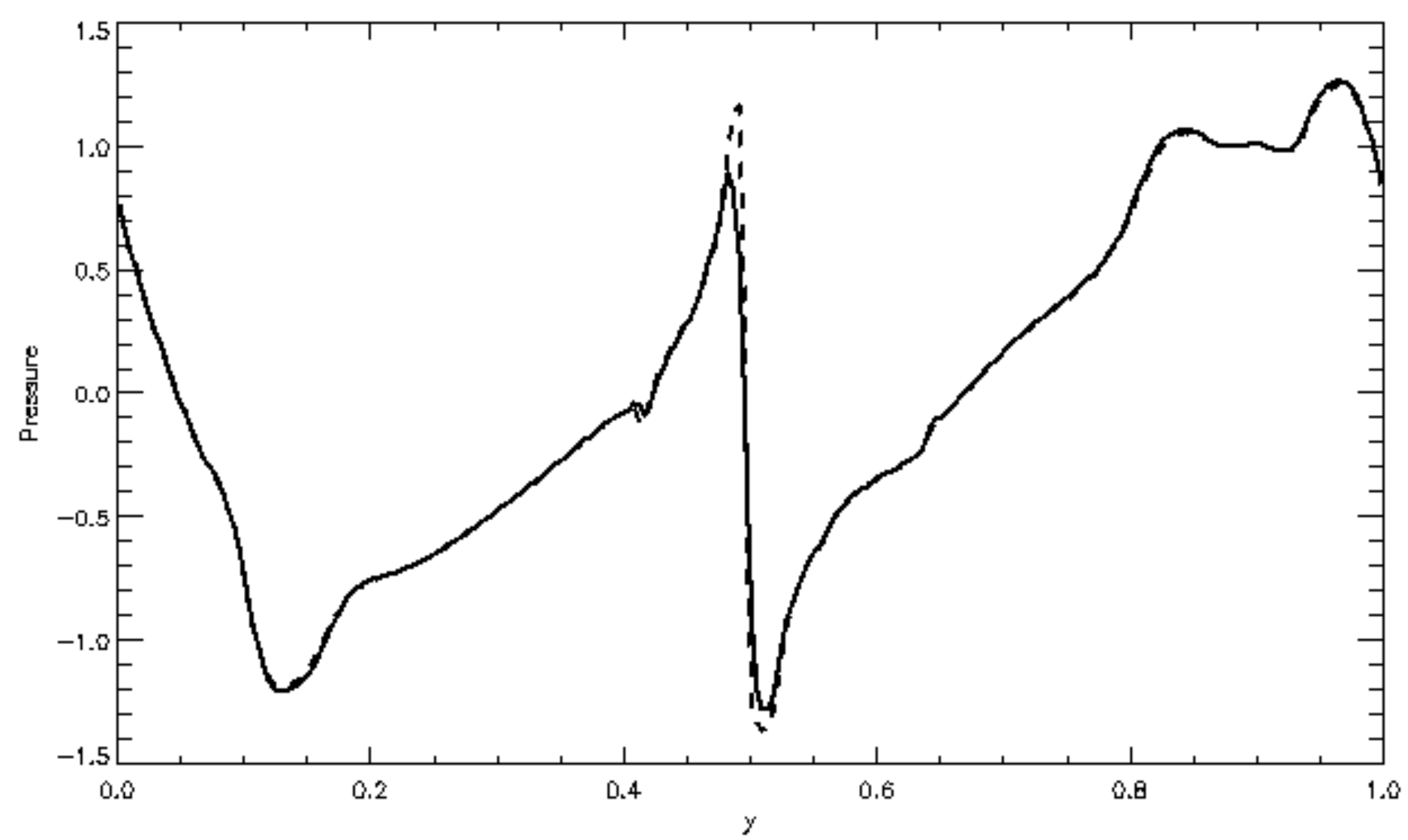}
 \caption{$B_x$ along a slice in the $y$-direction for the Orszag-Tang test. \textsf{HLL3R} (dashed) and \textsf{FLASH-HLLE} (solid). The result from \textsf{FLASH-Roe} was indistinguishable from \textsf{HLL3R}.}
  \label{fig:otcut}
\end{figure}

\begin{figure}
\begin{center}
\def\arraystretch{0.0}
\begin{tabular}{l}
\includegraphics[scale=.4]{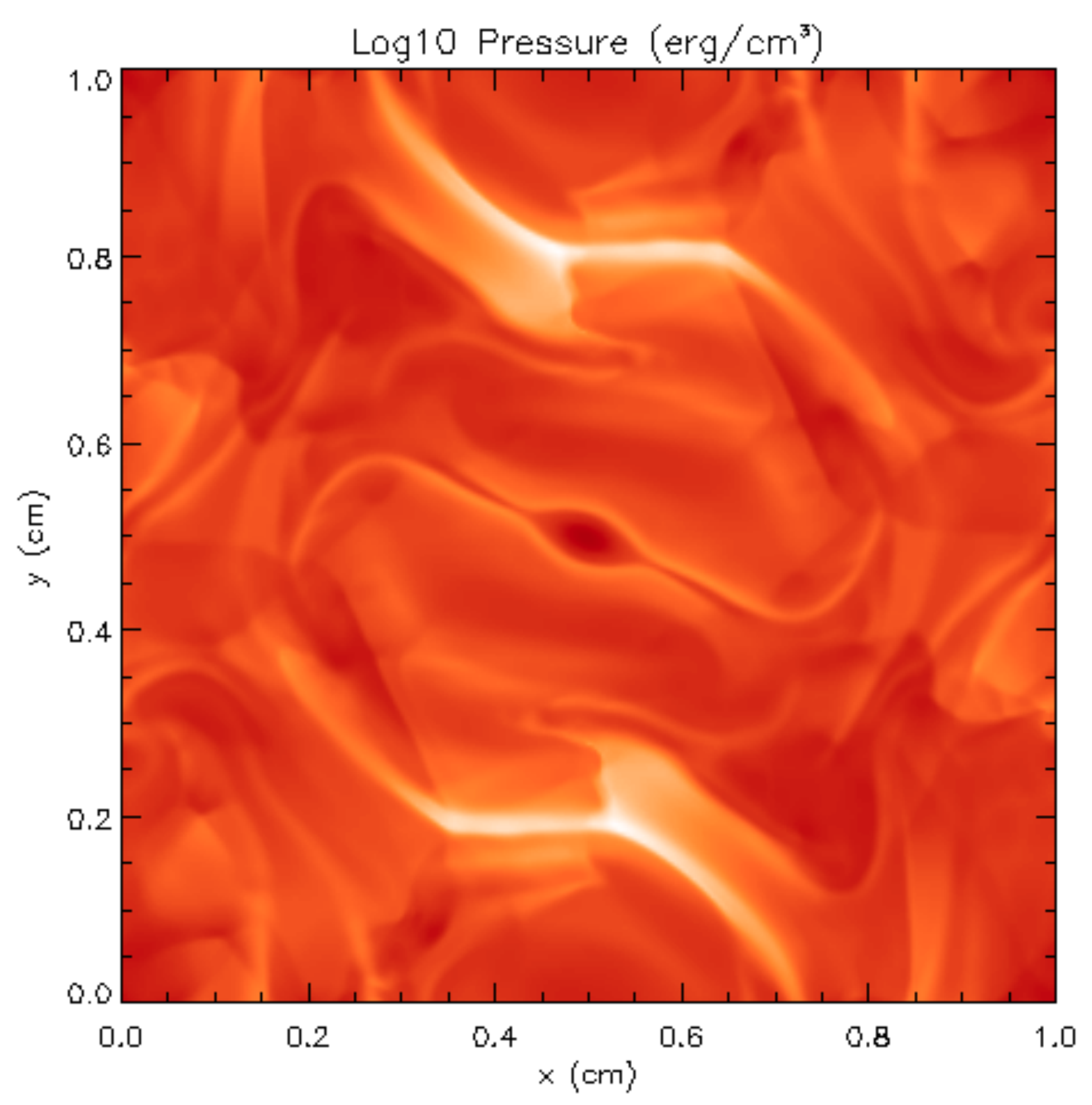} \\
\includegraphics[scale=.4]{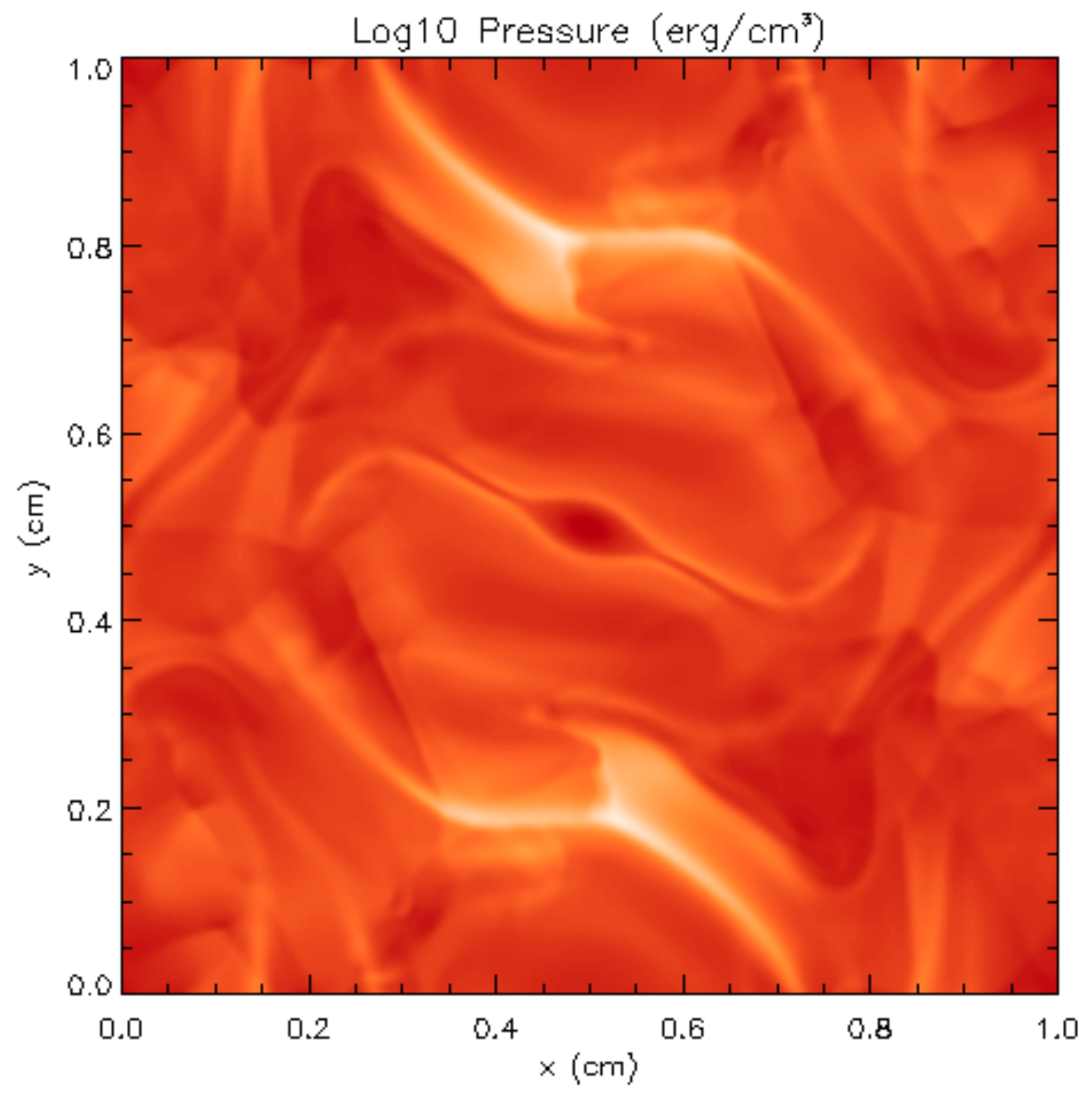} \\
\includegraphics[scale=.4]{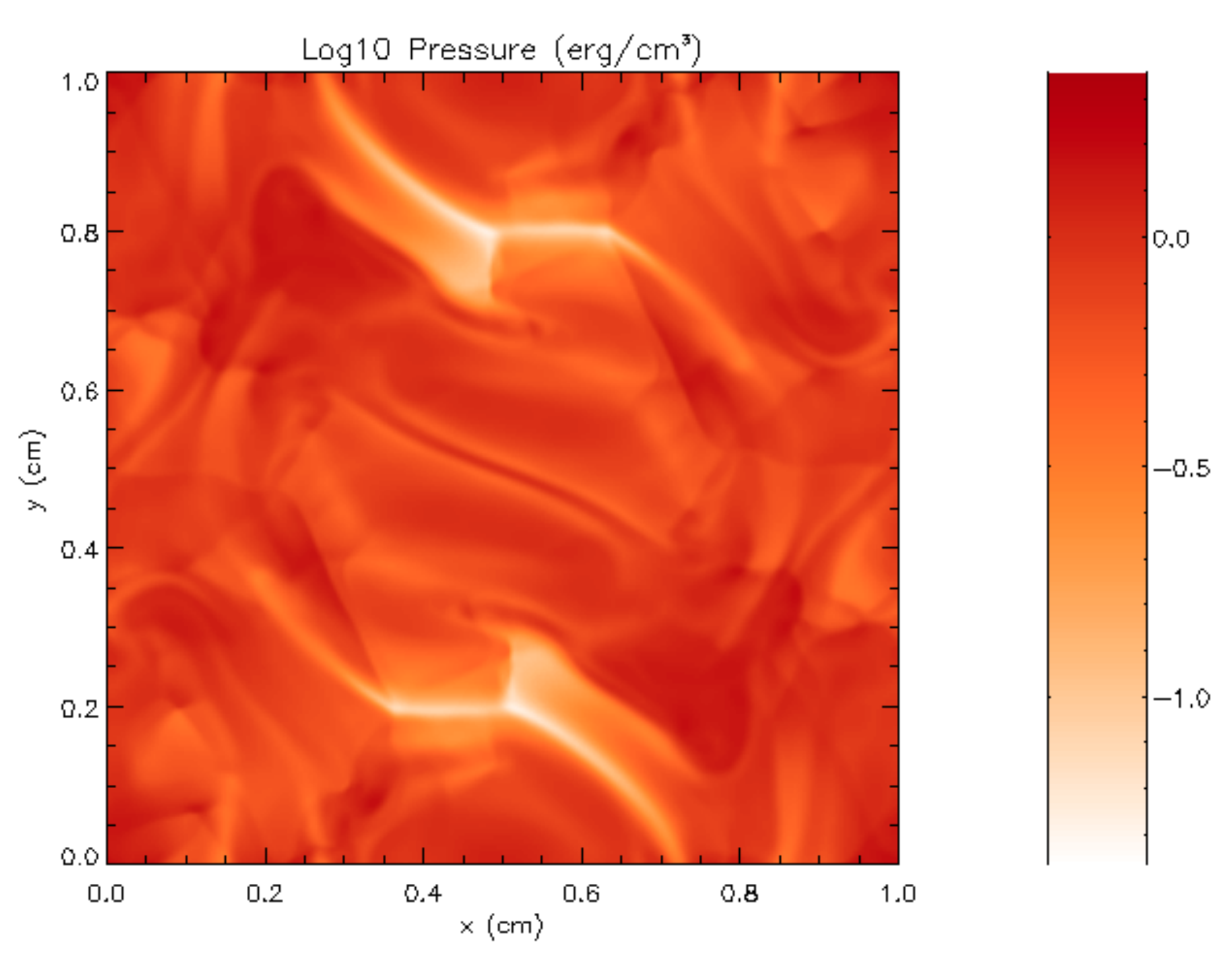}
\end{tabular}
\end{center}
\caption{Orszag-Tang test. Pressure at time $t=1$ with resolution $h=256^{-1}$. Top to bottom: \textsf{FLASH-Roe}, \textsf{HLL3R}, \textsf{FLASH-HLLE}.}
\label{fig:ot2}
\end{figure}

This is a standard test case for ideal MHD schemes. It consists of smooth velocity and magnetic field profiles going through shock formations, complicated wave interactions and eventually forming instabilities. The initial data are given by $\gamma=5/3$,
\begin{equation}
  \left(\rho, \bfu,\bfB,p\right)
  = \left(1, \sin(\pi y),
    \sin(\pi x),0,-\sin(\pi y),\sin(2 \pi x),0, 0.6\right). \label{eq:init2}
\end{equation}
At $t=0.5$ we observed only minor differences between \textsf{HLL3R} and \textsf{FLASH-Roe}, while \textsf{FLASH-HLLE} gave a smoother pressure peak along the central current sheet (see Fig.~\ref{fig:ot}). Plotting $B_x$ along the cut at $x=0.428$ (Fig.~\ref{fig:otcut}) shows that \textsf{FLASH-HLLE} gives a slightly more smeared-out current sheet. \textsf{HLL3R} and \textsf{FLASH-Roe} gave very similar results at $t=0.5$. At $t=1$ a magnetic island can be observed at the domain centre with \textsf{FLASH-Roe} and \textsf{HLL3R}, while with \textsf{FLASH-HLLE} this instability is suppressed by numerical viscosity (Fig.~\ref{fig:ot2}). The \textsf{HLL5R} was also tested and gave results very similar to \textsf{HLL3R} and \textsf{FLASH-Roe}.

We consider computational times in Table~\ref{tab:clock}. They are from a computation on two processors, in the form of total CPU time in seconds and in the percentage of that computation time spent in the sweep routine (which is the only differing one). This should be taken with a grain of salt, as these numbers vary a bit from time to time. We took the average of two runs. It appears that \textsf{HLL3R} gives about a 10\% speed-up compared to \textsf{FLASH-Roe}. \textsf{FLASH-HLLE} is also a little slower than \textsf{HLL3R}, which is due to the different reconstruction procedures. All in all we would expect \textsf{FLASH-HLLE} to be faster than \textsf{FLASH-Roe}, since only the computation of the numerical flux differs. The time to compute the numerical flux in \textsf{HLL3R} should be somewhere in between the time for \textsf{FLASH-Roe} and \textsf{FLASH-HLLE}, and in addition \textsf{HLL3R} has a simpler reconstruction and a slightly more elaborate Powell term than \textsf{FLASH-Roe} and \textsf{FLASH-HLLE}. The \textsf{HLL5R} scheme is the most efficient, however it requires slightly smaller time steps. The shorter time steps are not surprising, since the wave speeds of the HLL5R Riemann solver may be up to $10\%$ larger than with the corresponding 3-wave solver in some configurations; see \cite{BKW2}. Consequently, there would be a roughly $10\%$ reduction in time step size.

\begin{table}
   \centering
   \begin{tabular}{@{} lcccc @{}} 
      Scheme               & \textsf{HLL3R} &\textsf{HLL5R} & \textsf{FLASH-HLLE} & \textsf{FLASH-Roe} \\
      \hline
      $N$ time steps       & 477            & 539           & 476           & 482          \\
      Sweep time fraction  & 43.6\%         & 47.6\%        & 47.2\%        & 50.0\%       \\
      Average time / step  & 0.555$\,$s     & 0.520$\,$s    & 0.571$\,$s    & 0.604$\,$s
   \end{tabular}
   \caption{Computational times for the Orszag-Tang test.}
   \label{tab:clock}
\end{table}

\subsection{Cloud-shock interaction}

\begin{figure}
   \centering
   \includegraphics[scale=.43]{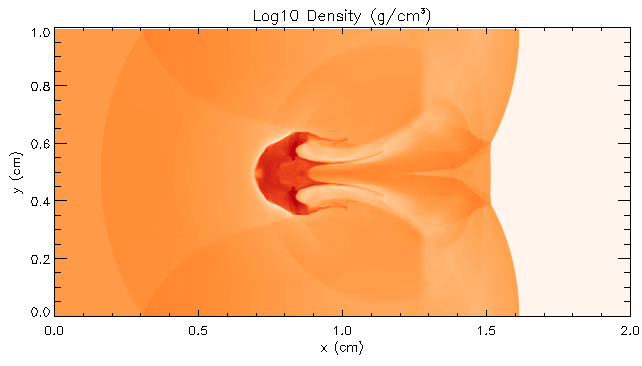} 
   \includegraphics[scale=.43]{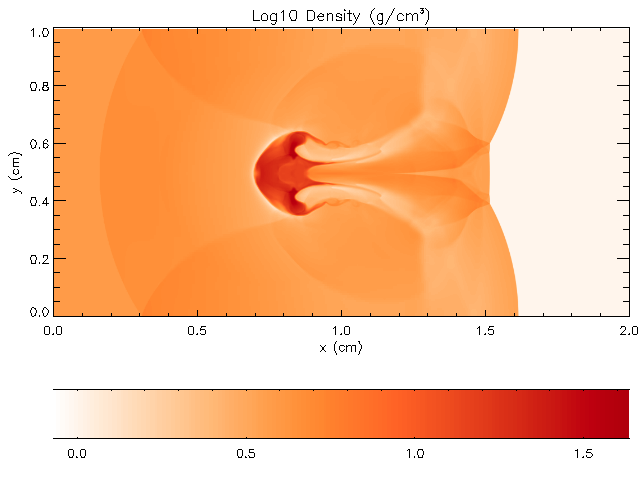}
   \caption{Cloud-shock interaction with 8 levels of refinement, corresponding to $1024\times2048$ cells. Grid and mass density. Top: \textsf{FLASH-Roe}, bottom: \textsf{HLL3R}}
   \label{fig:csamr}
\end{figure}

The initial conditions are the same as in \cite{FMMRWPowell}. The computational domain is $(x,y) \in
[0,2]\times[0,1]$ with artificial Neumann type boundary
conditions. The initial conditions consist of a shock moving to the
right, located at $x=0.05$, and a circular cloud of density $\rho=10$ and radius $r=0.15$ centred
at $(x,y) = (0.25,0.5)$. We took $\gamma=5/3$ and
\begin{align}
  \rho &=
  \begin{cases}
    3.86859 &\text{if $x<0.05$,}\\
    10.0    &\text{if $(x-0.25)^2+(y-0.5)^2<0.15^2$,}\\
    1.0     &\text{otherwise,}
  \end{cases}
\\
  \bfu &=
  \begin{cases}
    (11.2536,0,0)  &\text{if $x<0.05$,}\\
    (1.0,0,0)   &\text{otherwise,}
  \end{cases}
\\
  p &=
  \begin{cases}
    167.345  &\text{if $x<0.05$,}\\
    1.0      &\text{otherwise,}
  \end{cases}
\\
  \bfB &= 
  \begin{cases}
    (0,2.18261820,-2.18261820)  &\text{if $x<0.05$,}\\
    (0,0.56418958,0.56418958)   &\text{otherwise.}
  \end{cases}
\end{align}
Hence, the shock has a sonic Mach number around 11.8 and an Alfv{\'e}nic Mach number around 19.0. The solution becomes very complicated as the shock interacts with the cloud, and instabilities are observed along filaments of the cloud. We used these initial data to test the ability of the new scheme to handle adaptive mesh refinement. Eight levels of refinement were used with the highest resolution corresponding to a resolution of $1024\times2048$ grid cells. The refinement criterion was based on the mass density $\rho$ with the FLASH code parameters $\texttt{refine\_cutoff} = 0.75$ and $\texttt{derefine\_cutoff} = 0.25$. Mass density profiles from \textsf{HLL3R} and \textsf{FLASH-Roe} are shown in Figure~\ref{fig:csamr}. The front of the cloud looks more rounded with \textsf{HLL3R}, and the structures of the turbulent wake differ somewhat. Both \textsf{FLASH-Roe} and \textsf{HLL3R} agree qualitatively well, but since there is no analytic solution it is unclear how to make a quantitative assessment of the differences. \textsf{HLL5R} gave very similar results to \textsf{HLL3R} in this test. \textsf{FLASH-HLLE} roughly reproduced the shape of the cloud front, but gave a different structure in the turbulent wake.

\subsection{Kelvin-Helmholtz instability}
\begin{figure}
   \centering
   \includegraphics[scale=.43]{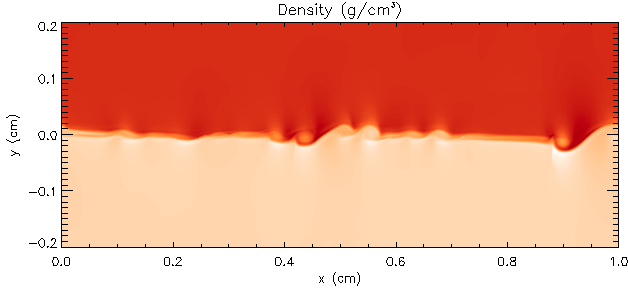}                \\
   \hspace{.015in} \includegraphics[scale=.43]{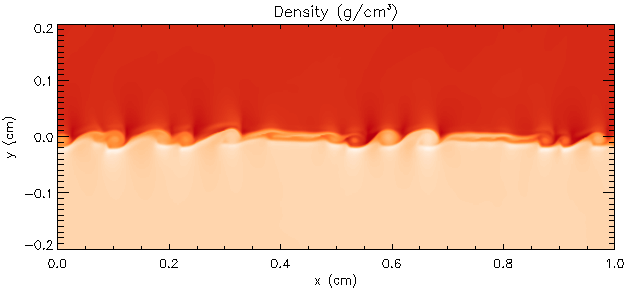} \\
   \hspace{.05in}\includegraphics[scale=.43]{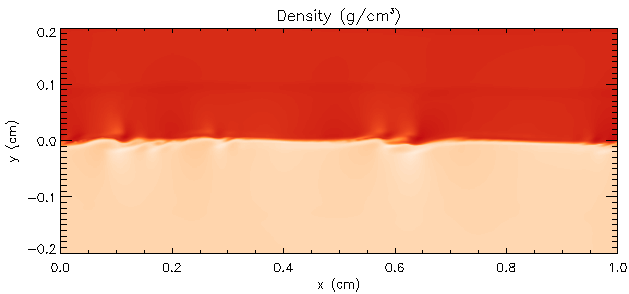}   \\
   \hspace{.12in}\includegraphics[scale=.43]{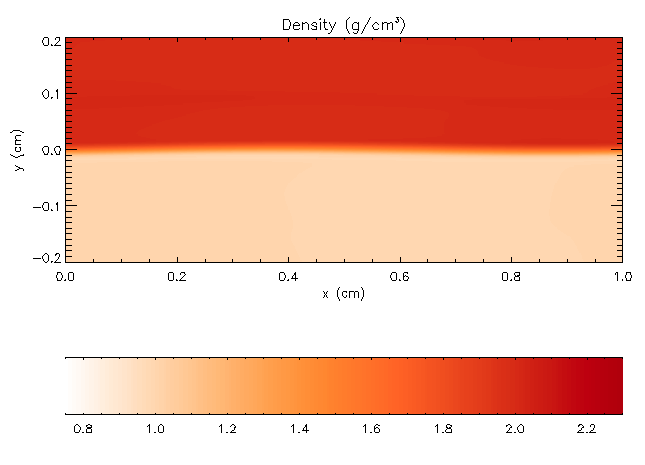}
\caption{Kelvin-Helmholtz instability at time $t=1$. Top to bottom: \textsf{FLASH-Roe}, \textsf{HLL5R}, \textsf{HLL3R}, \textsf{FLASH-HLLE}.}
\label{kh1}
\end{figure}

The Kelvin-Helmholtz instability is important as a model for the onset of turbulence. From a numerical point of view, we found it illustrative of some differences between the Riemann solvers. We consider a case where both the velocity and magnetic fields switch polarity along a straight line. The initial data are given by $p=3/5$, $\gamma=5/3$, $(v,w,B_y,B_z)=0$ and
\begin{align}
\rho=2, u=0.5, B_x=0.5, 
\end{align}
for $y>0$, and 
\begin{align}
\rho=1, u=-0.5, B_x=-0.5, 
\end{align}
for $y<0$. We perturb the velocity in the $y$-direction by
\beq
v=0.0125\,\sin(2\pi x)\,\exp\left[-100(y-0.5)^2\right] .
\label{vperturb}
\eeq
At $y=-0.5$ and $y=0.5$ we impose reflecting boundary conditions. Hence, we consider a small perturbation to a tangential discontinuity in both $\bfu$ and $\bfB$ along the line $y=0$. Initially the relative sonic Mach number is $1$, the magnetosonic Mach number is $2/\sqrt{5}$, while plasma $\beta=8\gamma$. We used a resolution of $512^2$ cells. In numerical experiments of this type, instability growth tends to occur at the small scales, and therefore depend on numerical viscosity. In \cite{KSW2007}, this was reflected in a strong dependence between grid resolution and instability growth rate. Mass density profiles at time $t=1$ are shown in Figure~\ref{kh1}.
The higher numerical viscosity of the HLLE solver results in \textsf{FLASH-HLLE} suppressing all instabilities at this time. \textsf{HLL3R} yields a sharper shear-discontinuity than \textsf{FLASH-HLLE}, but the roll-ups still did not develop. In contrast, \textsf{HLL5R} and \textsf{FLASH-Roe} produce well-developed Kelvin-Helmholtz roll-ups.

The different behaviour of the schemes can be explained by the approximate Riemann solvers. By the remarks at the end of section~\ref{first_order}, Roe and HLL5R exactly resolve the unperturbed initial configuration, while HLL3R smears out the velocity shear slightly. HLLE smears out the material and magnetic contact discontinuities of the initial configuration, with the result that the instability is entirely suppressed at early times. The presence of small perturbations to grid-aligned stationary discontinuities makes this test case particularly sensitive to these properties of the Riemann solvers. In particular, we see the effect of exactly resolving the discontinuities, i.e., having a scheme that is 'well-balanced' with respect to the unperturbed state. In non-stationary or non-grid-aligned cases, none of the schemes have this well-balancing property, hence we expect differences to be less pronounced.

\begin{figure}
   \includegraphics[scale=.38]{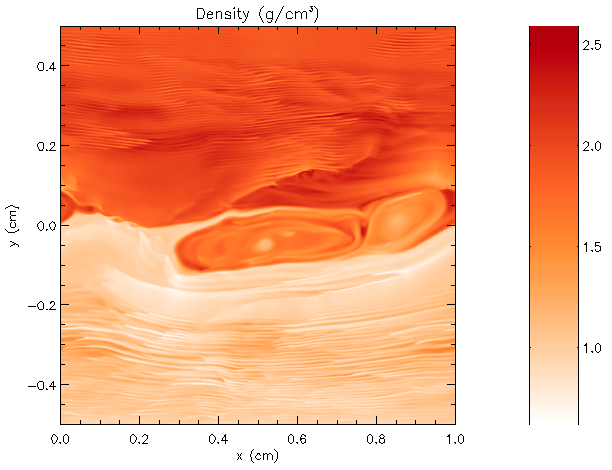}
   \includegraphics[scale=.38]{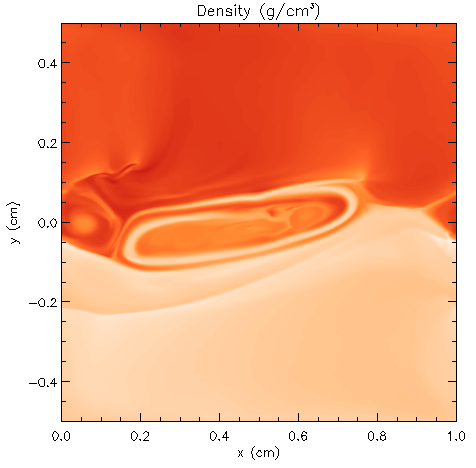}
   \caption{Kelvin-Helmholtz at time $t=9$, \textsf{FLASH-Roe} (left) and \textsf{HLL5R} (right)}.
   \label{kh9}
\end{figure}

At late times of the instability development, it is harder to assess the performance of the schemes. Results from \textsf{FLASH-Roe} and \textsf{HLL5R} are shown in Figure~\ref{kh9}. Both schemes give a similar structure of two larger vortices. \textsf{FLASH-Roe} produces small-scale waves in the surrounding medium, which are possibly a numerical artefact. One possible source of spurious features in \textsf{FLASH-Roe} is the central discretisation of the Powell source term. Such errors have been pointed out in \cite{FMMRWPowell}, \cite{fkrsid1}, \cite{KW2010} and \cite{W1}.

\begin{figure}
   \includegraphics[scale=.4]{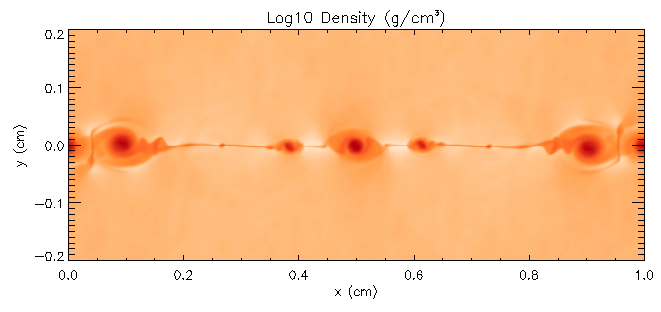}
   \includegraphics[scale=.4]{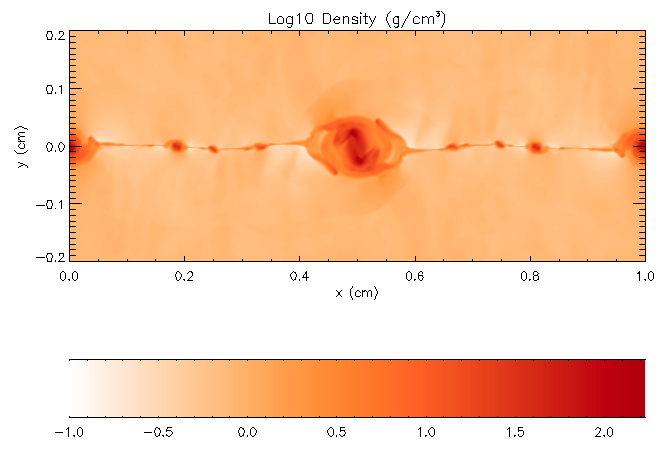}
   \caption{High sonic Mach number Kelvin-Helmholtz test for isothermal gas at time $t=1$. Top: \textsf{HLL3R}. Bottom: \textsf{HLL5R}.}
   \label{fig:kh_isoT}
\end{figure}

Finally, in order to demonstrate the robustness of \textsf{HLL3R} and \textsf{HLL5R}, we tested an isothermal case with a higher sonic Mach number. The sound speed was set to $0.1$, and the initial data were $(v,w,B_y,B_z)=0$ everywhere,
\begin{align*}
\rho=1, u=0.5, B_x=2, 
\end{align*}
for $y>0$, and 
\begin{align*}
\rho=1, u=-0.5, B_x=-2, 
\end{align*}
for $y<0$. The perturbation was again given by~\eqref{vperturb}. Hence, the relative sonic Mach number was 10, and the relative Alfv{\'e}nic Mach number was $1/\sqrt{2}$. Resolution was uniform with $512^2$ grid cells. The standard schemes, \textsf{FLASH-Roe} and \textsf{FLASH-HLLE} failed on this case due to unphysical states, but the new schemes, \textsf{HLL3R} and \textsf{HLL5R} produced physical results. Density profiles from \textsf{HLL3R} and \textsf{HLL5R} are shown in Fig.~\ref{fig:kh_isoT}. The resulting instability contains vortex-like structures with complicated internal features. Due to the chaotic nature of these solutions, we cannot evaluate the differences seen between the schemes.

\clearpage
\section{Supersonic MHD turbulence}
In this section we investigate the performance of three of the MHD schemes, \textsf{FLASH-Roe}, \textsf{HLL3R} and \textsf{HLL5R} on the modelling of supersonic MHD turbulence. Testing the schemes in the regime of highly compressible turbulent flows is of particular interest for astrophysical applications ranging from molecular cloud formation on galactic scales down to scales of individual star-forming accretion disks. These disks can launch powerful supersonic jets, driven by magneto-centrifugal forces \cite{BlandfordPayne1982} and magnetic towers \cite{LyndenBell2003}. Here however, we concentrate on intermediate scales of molecular cloud evolution, which is largely determined by the interplay of supersonic turbulence and magnetic fields. Understanding the nature and origin of supersonic MHD turbulence is considered to be the key to understanding star formation in molecular clouds \cite{MacLowKlessen2004,ElmegreenScalo2004,ScaloElmegreen2004,McKeeOstriker2007}.

\subsection{Numerical setup and post-processing}
We start with a short description of the type of setup often adopted in numerical models of molecular cloud turbulence. It is called turbulence-in-a-box, because a three-dimensional box with periodic boundary conditions and random forcing is used to excite turbulence. We use an isothermal equation of state throughout all turbulence-in-a-box runs, i.e., the pressure, $p_\mathrm{th}$, is related to the density by $p_\mathrm{th}=\cs^2\rho$ with the constant sound speed, $\cs$. For these tests we used an initially uniform density, $\langle\rho\rangle=1.93\times10^{-21}\,\g\,\cm^{-3}$, corresponding to a mean particle number density of $\langle n\rangle=\langle\rho\rangle/(\mu m_\mathrm{p})\approx500\,\cm^{-3}$, with the mean molecular weight, $\mu=2.3$ and the proton mass, $m_\mathrm{p}$, and a box size of $L^3=(4\,\pc)^3$. The sound speed for the typical gas temperatures in molecular clouds (about $11\,\K$) is roughly $\cs=0.2\,\km\,\s^{-1}$. For driving the turbulence we use the same forcing module as used in \cite{FederrathKlessenSchmidt2008,FederrathKlessenSchmidt2009,SchmidtEtAl2009,PriceFederrath2010}, which is based on the stochastic Ornstein-Uhlenbeck process \cite{EswaranPope1988}. The spectrum of the forcing amplitude is a paraboloid in Fourier space in a small wavenumber range $1<|\vect{k}|<3$, peaking at $|\vect{k}|=2$. This corresponds to injecting most of the kinetic energy on scales of half of the box size. The turbulence forcing procedure is explained in detail in \cite{FederrathDuvalKlessenSchmidtMacLow2010}. We use a solenoidal (divergence-free) forcing throughout this study unless noted otherwise. For the comparison of the different MHD schemes the forcing amplitude was set such that the turbulence reaches a root-mean-squared (RMS) Mach number, $\Ma\approx2$ in the fully developed, statistical steady regime. The resolution was fixed at $256^3$ grid cells. We also consider a highly supersonic case with $\Ma\approx10$ in section~\ref{sec:turb_resol}, for which we investigate the resolution dependence by comparing runs with $128^3$, $256^3$ and $512^3$ grid zones. This study, however, could only be followed with the new types of solvers presented here (HLL3R and HLL5R), because the Roe solver is numerically unstable for highly supersonic turbulence. As shown in previous studies \cite{FederrathKlessenSchmidt2009,SchmidtEtAl2009,FederrathDuvalKlessenSchmidtMacLow2010} turbulence is fully developed after about two large-scale eddy turnover times, $2T$, where one large-scale eddy turnover time is defined as $T=L/(2\Ma\cs)$. The RMS velocity is thus defined as $V=L/(2T)=\Ma c_s$.

In order to test the MHD schemes in a strongly magnetised case, we set the initial uniform magnetic field strength to $B_0=8.8\times10^6\,\G$ in the $z$-direction, which gives an initial plasma beta, $\beta=p_{th}/\pmagzero\approx0.25$, with the magnetic pressure, $\pmagzero=B_0^2/(8\pi)$. No initial fluctuations of the magnetic field were added, such that the initial RMS magnetic field is equal to $B_0$. However, the turbulence twists, stretches, and folds the initially uniform magnetic field, such that the RMS field strength increases due to turbulent dynamo amplification until saturation (see the review by Brandenburg and Subramanian 2005 \cite{BrandenburgSubramanian2005};  and section \ref{dynamo}). The Alfv\'{e}nic Mach number is $\MA=V/v_\mathrm{A}=\Ma\sqrt{\beta/2}\approx0.7$, i.e., we consider slightly sub-Alfv\'{e}nic turbulence. The \textsf{HLL3R} scheme, however, was used in both the highly sub-Alfv\'{e}nic ($\MA\ll1$) and the highly super-Alfv\'{e}nic ($\MA\gg1$) turbulent regimes by \cite{BruntFederrathPrice2010a,BruntFederrathPrice2010b}, who modelled turbulent flows with $\beta=0.01\dots10$, $\MA=0.4\dots50$ and $\Ma=2\dots20$.

All results were averaged over one eddy turnover time, $T$, in the regime of fully developed turbulence, $t>2\,T$ \cite{FederrathDuvalKlessenSchmidtMacLow2010}. Since we produced output files every $0.1\,T$, we computed Fourier spectra and probability distribution functions for each individual snapshot and averaged afterwards over 11 snapshots for $2.3\leq t/T\leq3.3$. The averaging procedure allows us to study the temporal fluctuations of all statistical measures and to improve the statistical significance of the similarities and the differences seen between the different schemes. This is an important advantage over comparing instantaneous snapshots, because it must be kept in mind that individual snapshots can be different due to intermittent turbulent fluctuations. These fluctuations can either make the results look very similar or very different between the different schemes, if one compares instantaneous snapshots only. Thus, a meaningful comparison can only be made by time averaging the results and considering the statistically average behaviour of the schemes (see also, \cite{PriceFederrath2010}).

\subsection{Time evolution}
\begin{figure}[tbp]
\centering
\includegraphics[width=0.6\linewidth]{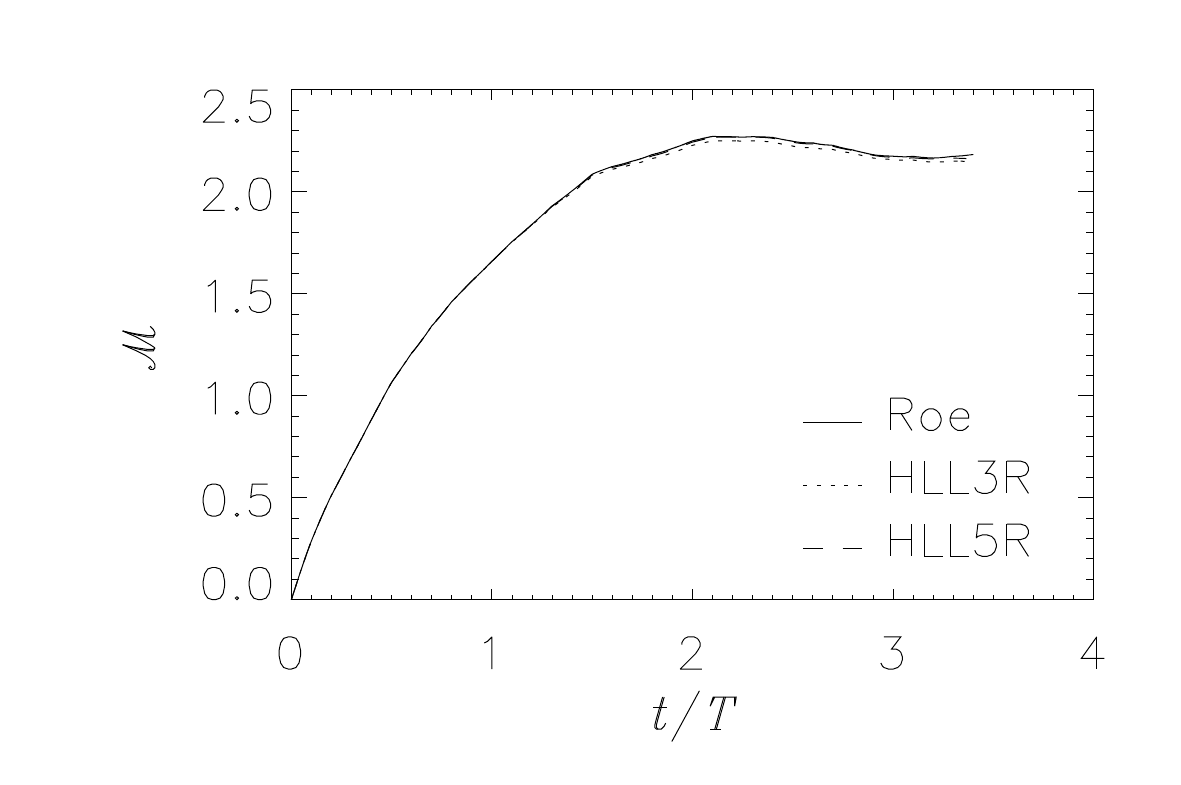}
\caption{Time evolution of the RMS Mach number, $\Ma$ for \textsf{FLASH-Roe}, \textsf{HLL3R} and \textsf{HLL5R}, shown as solid, dotted and dashed lines, respectively. Turbulence is fully developed for $t\gtrsim2T$, where $T$ is the large-scale eddy turnover time.}
\label{fig:turb_machevol}
\end{figure}

Figure~\ref{fig:turb_machevol} shows the time evolution of the RMS Mach number, $\Ma=\sqrt{\langle\vect{u}^2\rangle}/\cs$. The magnetised fluid is accelerated from rest by the random stochastic forcing as discussed in the previous section. It reaches a statistical steady state at around $t\approx2\,T$, such that turbulence is fully developed for $t\gtrsim2\,T$ (see also, \cite{FederrathKlessenSchmidt2009,FederrathDuvalKlessenSchmidtMacLow2010}). We can thus safely use time-averaging for $2.3\leq t/T\leq3.3$ of the Fourier spectra discussed in the next section.

\subsection{Fourier spectra}
\begin{figure}[tbp]
\centering
\includegraphics[width=0.6\linewidth]{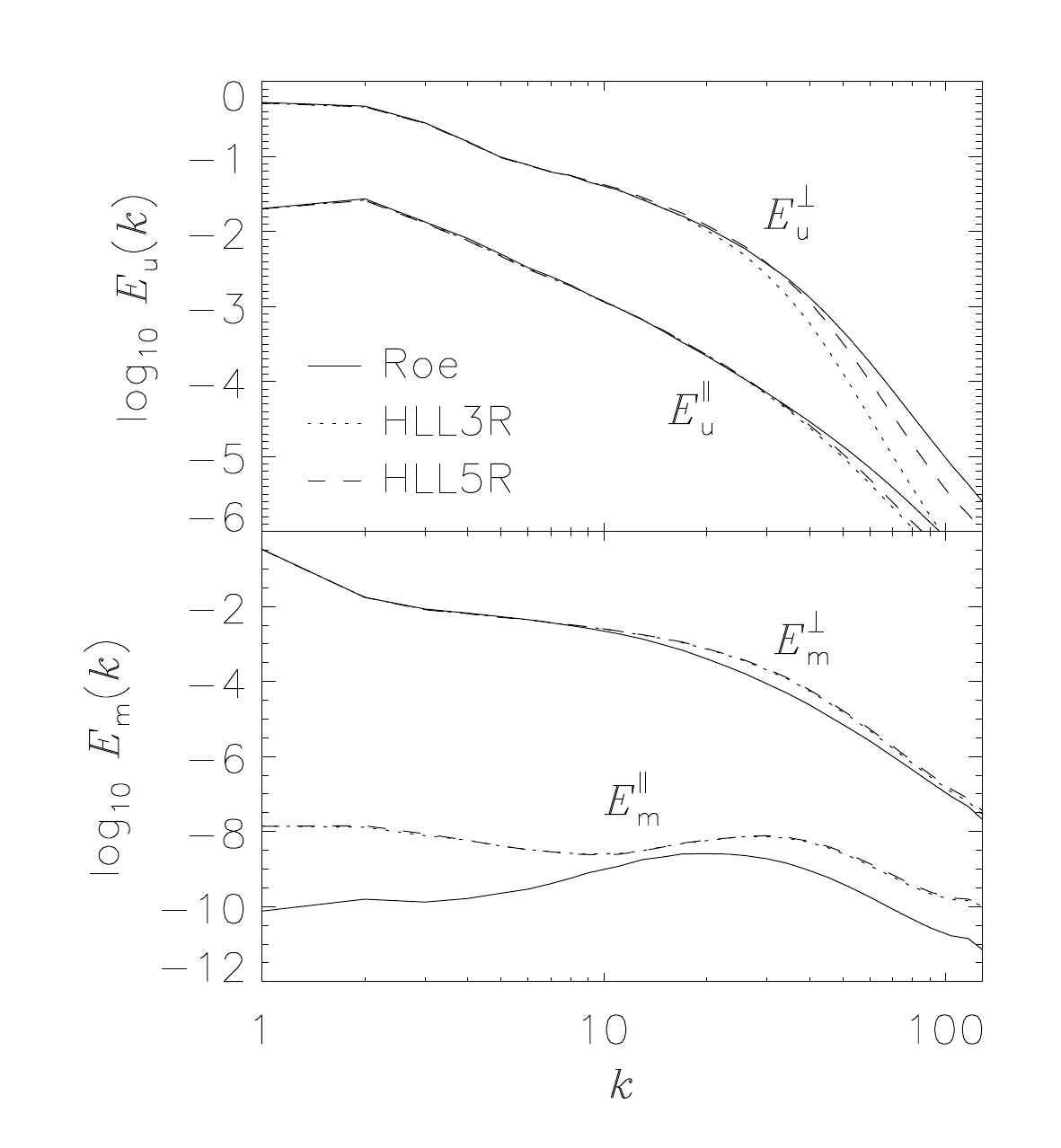}
\caption{Comparison of the Fourier spectra of the turbulent velocity (\emph{top panel}) and the turbulent magnetic field (\emph{bottom panel}) for the \textsf{FLASH-Roe}, \textsf{HLL3R} and \textsf{HLL5R}, shown as solid, dotted and dashed lines, respectively. The numerical resolution was $256^3$ grid cells. Both the velocity and the magnetic field spectra were decomposed into their rotational and compressible parts, $E^\perp$ and $E^\parallel$, respectively. The temporal variations of the spectra are on the order of three times the line width in this plot.}
\label{fig:turb_spect}
\end{figure}

As a typical measure used in turbulence analyses (e.g., \cite{Frisch1995,KSW2007,KitsionasEtAl2009}), we show Fourier spectra of the velocity and magnetic fields in Figure~\ref{fig:turb_spect}, decomposed into their rotational part, $E^\perp$ and their longitudinal part, $E^\parallel$. For the velocity spectra (top panel), $E_\mathrm{u}^\perp\propto|\vect{u}_\perp|^2$ is a measure for the rotational motions of the turbulent flow, while $E_\mathrm{u}^\parallel\propto|\vect{u}_\parallel|^2$ is a measure for the compressions induced by shocks. Since here we drive the turbulence with a solenoidal random force \cite{FederrathDuvalKlessenSchmidtMacLow2010}, and since the RMS Mach number is only slightly supersonic ($\Ma\approx2$) the bulk of the turbulent motions is in the rotational part, $E_\mathrm{u}^\perp$. From the velocity spectra we conclude that \textsf{FLASH-Roe} is the least dissipative scheme of the three schemes tested. The \textsf{HLL3R} is slightly more dissipative, which is seen from the faster loss of kinetic energy at wavenumbers $k\gtrsim20$. The \textsf{HLL5R} is in between the \textsf{FLASH-Roe} and the \textsf{HLL3R} spectrum. Given that the limited numerical resolution does not allow for an accurate determination of the turbulence spectrum in the inertial range, the differences for $k\gtrsim20$ are acceptable (see also section~\ref{sec:turb_resol} below). \cite{KitsionasEtAl2009} and \cite{FederrathDuvalKlessenSchmidtMacLow2010} find that about 30 grid cells are necessary to resolve the turbulence on the smallest scales. This means that the turbulence on all wavenumbers $k\gtrsim N/30\approx8$ may be affected by the limited resolution (here $N=256$), which means that the differences between the different schemes for $k\gtrsim20$ can be safely ignored.

The magnetic field spectra (Figure~\ref{fig:turb_spect}, bottom panel) display small differences for $k\gtrsim10$. As for the velocity spectra the differences occur on scales smaller than the scales that are affected by numerical resolution, which means that these differences are negligible in applications of supersonic MHD turbulence. The rotational part of the magnetic field, $E_\mathrm{m}^\perp$, shows slightly less dissipation for the \textsf{HLL3R} and \textsf{HLL5R} compared \textsf{FLASH-Roe}. The longitudinal part, $E_\mathrm{m}^\parallel$, is a measure for $(\nabla\cdot\vect{B})^2$. The \textsf{FLASH-Roe} scheme keeps the divergence of the magnetic field smaller by an order of magnitude compared to \textsf{HLL3R} and the \textsf{HLL5R} on all scales. However, all schemes maintain the $\nabla\cdot\vect{B}$ constraint within acceptable values. The ratio $\int{E_\mathrm{m}^\parallel\,\mathrm{d}k}/\int{(E_\mathrm{m}^\parallel+E_\mathrm{m}^\perp)\,\mathrm{d}k} < 5.5\times10^{-7}$ for all times, hence numerical $\nabla\cdot\vect{B}$ effects did not have any significant dynamical influence.

\subsection{Probability distribution functions}
\begin{figure}[tbp]
\centering
\includegraphics[width=0.495\linewidth]{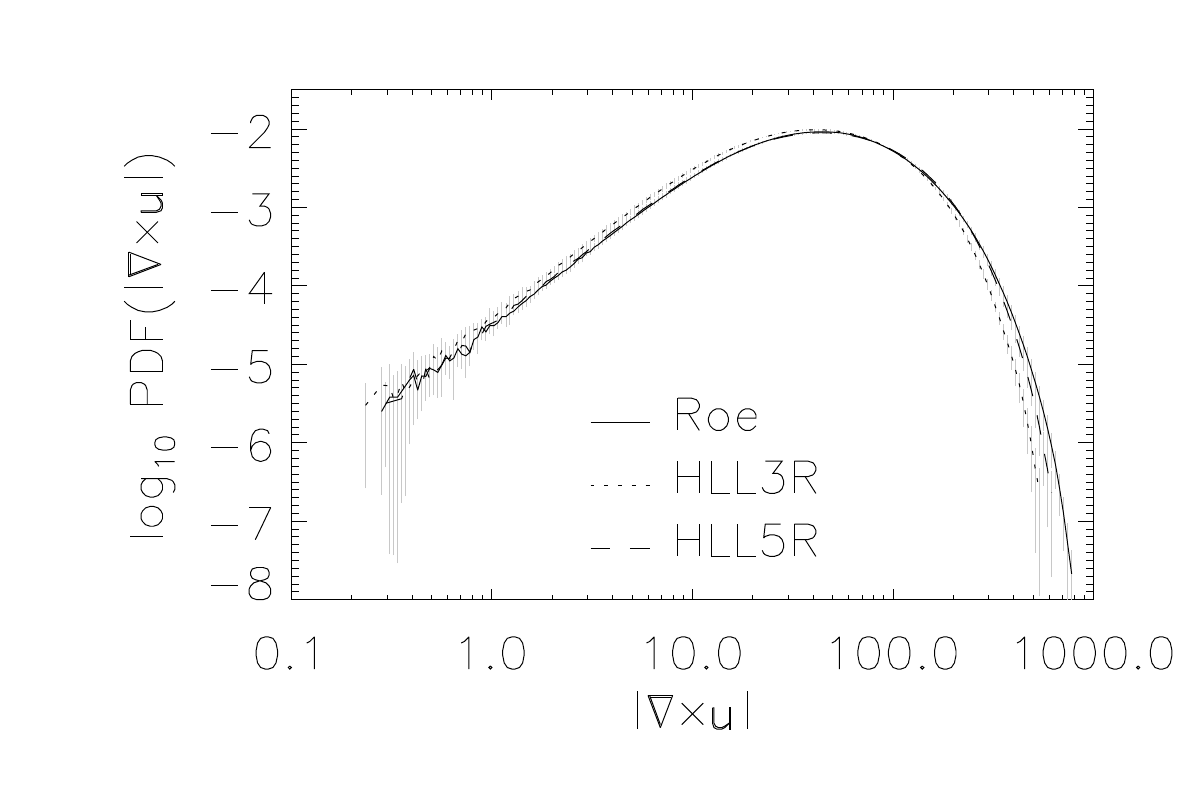}
\includegraphics[width=0.495\linewidth]{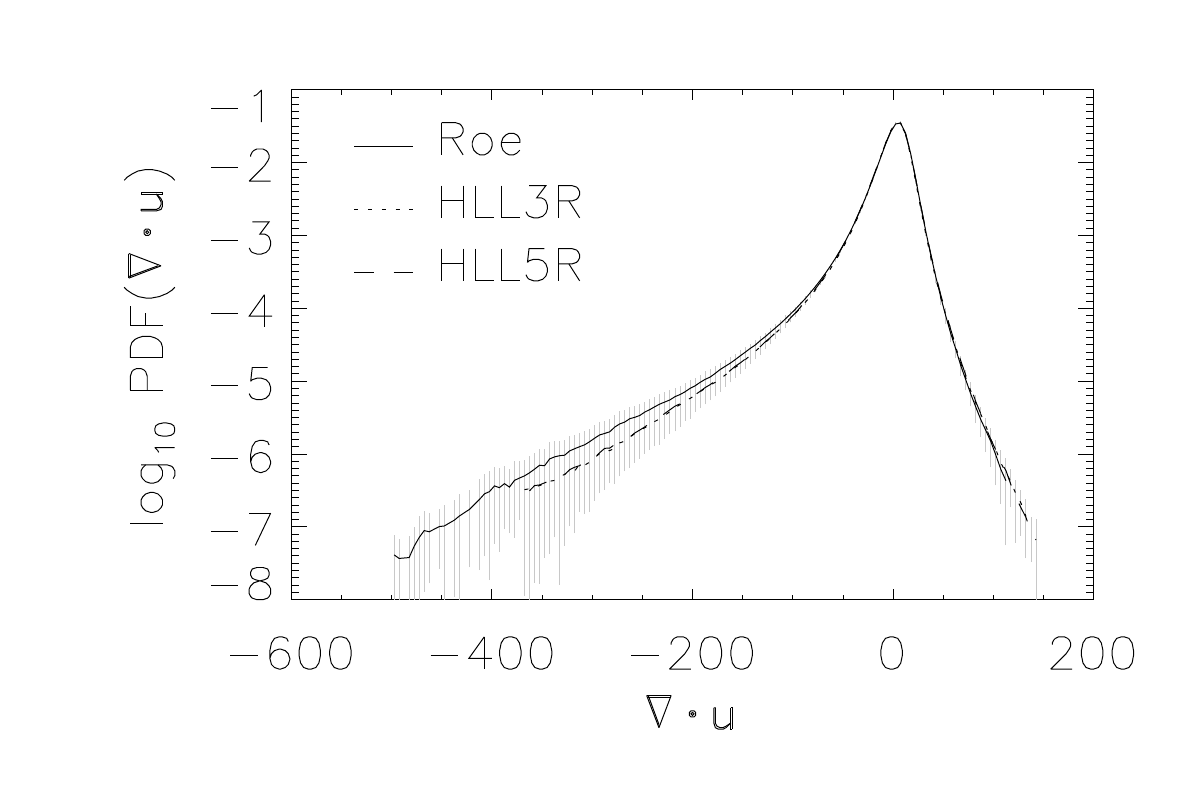}
\includegraphics[width=0.495\linewidth]{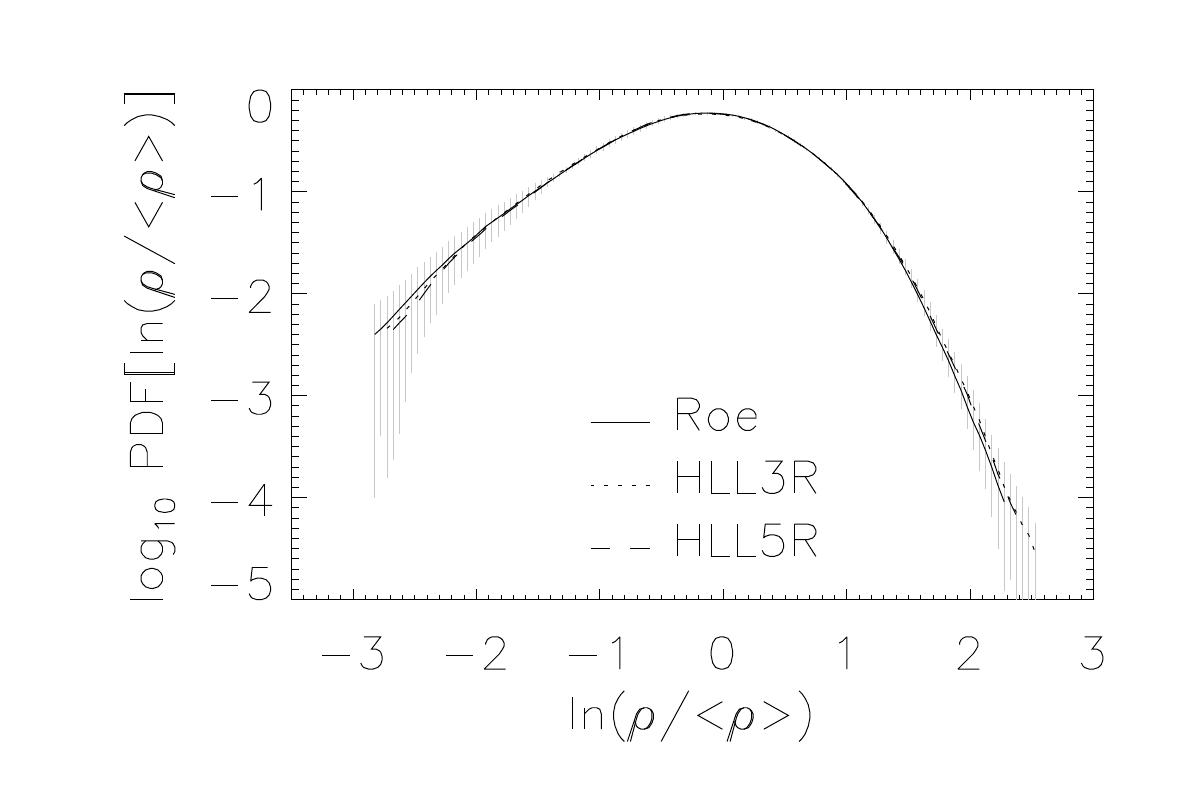}
\includegraphics[width=0.495\linewidth]{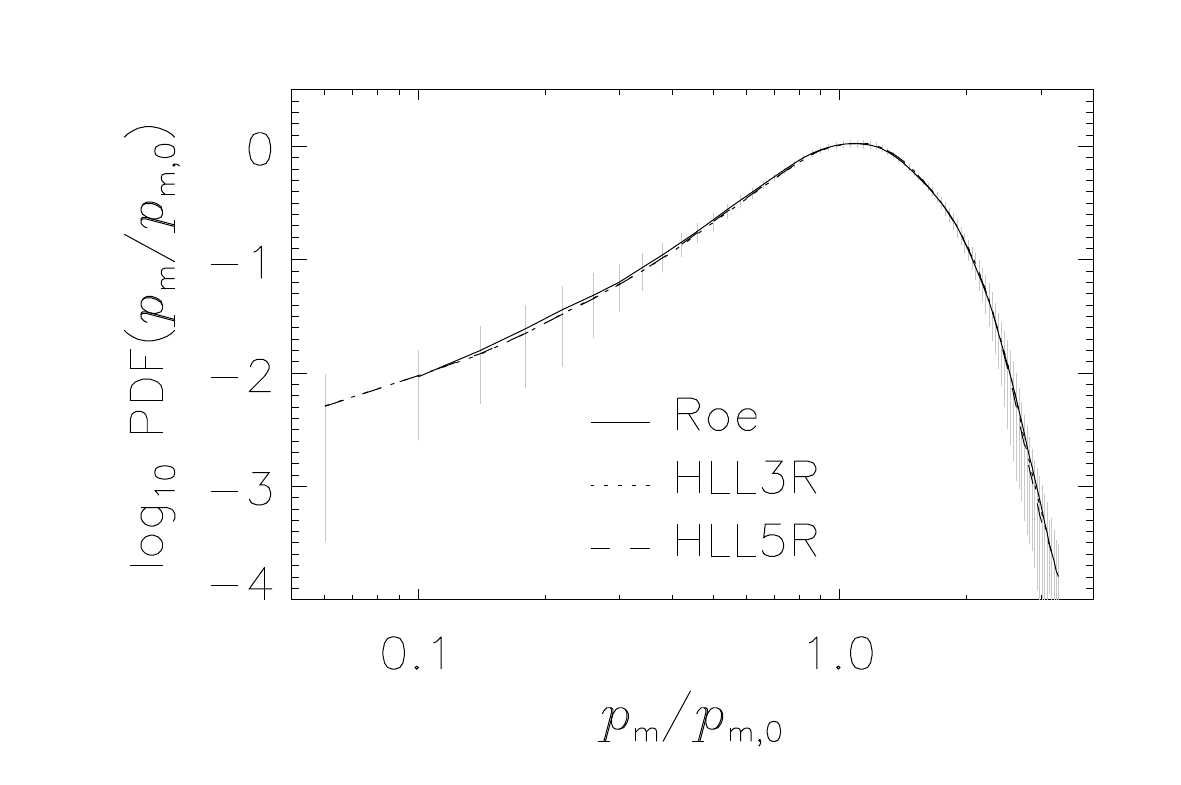}
\caption{Probability distribution functions of the vorticity, $|\nabla\times\vect{u}|$ (\emph{top left}), the divergence of the velocity, $\nabla\cdot\vect{u}$ (\emph{top right}), the logarithmic density, $s=\ln(\rho/\langle\rho\rangle)$ (\emph{bottom left}), and the magnetic pressure, $\pmag/\pmagzero=B^2/B_0^2$ (\emph{bottom right}), for \textsf{FLASH-Roe}, \textsf{HLL3R} and \textsf{HLL5R}, shown as solid, dotted and dashed lines, respectively. The numerical resolution was $256^3$ grid cells. The grey hatched regions indicate the 1-$\sigma$ temporal fluctuations of the distributions.}
\label{fig:turb_pdfs}
\end{figure}

To further investigate the dissipative properties of the different schemes we show probability distribution functions (PDFs) of the vorticity, $|\nabla\times\vect{u}|$, in Figure~\ref{fig:turb_pdfs} (top left panel). The vorticity is a measure of rotation in the turbulent flow. Stronger numerical dissipation leads to a faster decay of small-scale eddies. Figure~\ref{fig:turb_pdfs} (top left panel) shows that the most dissipative among the three schemes tested, \textsf{HLL3R} produces smaller values of the mean vorticity by about 10\% compared to the \textsf{FLASH-Roe}. The mean vorticity achieved with \textsf{HLL5R} is equivalent to \textsf{FLASH-Roe}. The RMS values of the vorticity for \textsf{HLL3R} and \textsf{HLL5R} are 11\% and 1\%, respectively smaller than for \textsf{FLASH-Roe}. This confirms the expected ranking of the schemes in terms of their numerical dissipation from the previous section: the \textsf{HLL3R} is the most dissipative, while the \textsf{HLL5R} is only marginally more dissipative than \textsf{FLASH-Roe}. 

The PDF of the divergence of the velocity field, $\nabla\cdot\vect{u}$ is shown in the top right panel of Figure~\ref{fig:turb_pdfs}. The asymmetry between rarefaction ($\nabla\cdot\vect{u}>0$) and compression ($\nabla\cdot\vect{u}<0$) is typical for compressible, supersonic turbulence (see, e.g., \cite{PassotVazquez1998,SchmidtEtAl2009}). \textsf{FLASH-Roe} produces slightly more zones with higher compression, which can be attributed to its lower numerical dissipation compared to \textsf{HLL3R} and \textsf{HLL5R} (cf., the Fourier spectra discussed in the previous subsection). However, all schemes agree to within the error bars, indicating strong temporal variations of compressed regions due to the intermittent nature of the shocks.

The PDF of the gas density is an essential ingredient for analytic models of star formation (e.g., \cite{PadoanNordlund2002,KrumholzMcKee2005,HennebelleChabrier2008}). All those models need the density PDF to estimate the mass fraction above a given density threshold by integrating the PDF. Thus, many studies have focused on investigating the density PDF in a supersonic turbulent medium (e.g., \cite{Vazquez1994,PadoanNordlundJones1997,PassotVazquez1998,Klessen2000,OstrikerStoneGammie2001,LiKlessenMacLow2003,JappsenEtAl2005,LemasterStone2008,FederrathKlessenSchmidt2008}). Figure~\ref{fig:turb_pdfs} (bottom left panel) shows the PDF of the logarithmic density, $s=\ln(\rho/\langle\rho\rangle)$, where $\langle\rho\rangle$ is the mean volume density. In this transformation of the density, the PDF follows closely a Gaussian distribution, i.e., a log-normal distribution in $\rho$ with a standard deviation of $\sigma_s^2=\ln\left(1+b^2\Ma^2\right)$ (see, e.g., \cite{FederrathDuvalKlessenSchmidtMacLow2010}). From this expression we derive a turbulence forcing parameter of $b=0.35\pm0.05$ for the given RMS Mach number, $\Ma=2.3\pm0.2$ and $\sigma_s=0.71\pm0.04$ for all three schemes. This is in very good agreement with the expected forcing parameter for purely solenoidal forcing as applied in this study, $b=1/3$ (see~\cite{FederrathKlessenSchmidt2008,FederrathDuvalKlessenSchmidtMacLow2010}).

Finally, we show the PDF of the magnetic pressure, $\pmag=B^2/(8\pi)$ in units of the initial magnetic pressure, $\pmagzero$, in Figure~\ref{fig:turb_pdfs} (bottom right panel). No significant differences are seen between the three different schemes. Thus, the magnetic pressure distribution is quite robust.

\subsection{Computational performance of the MHD schemes}
All runs with the three different schemes, \textsf{FLASH-Roe}, \textsf{HLL3R} and \textsf{HLL5R} were performed on the same machine, the HLRB II at the Leibniz Supercomputing Center in Munich in a mode of parallel computation (MPI) with 64 CPUs. In Table~\ref{tab:turb_cputime} we list the wall clock time per step, the total number of steps, the total amount of CPU hours spent, and the CFL number used for each scheme in our driven MHD turbulence experiments with a numerical resolution of $256^3$ grid cells. Only for \textsf{FLASH-Roe} we had to use a CFL number of 0.2 instead of 0.8 (both \textsf{HLL3R} and \textsf{HLL5R}), because it became unstable for CFL numbers $>0.2$ and crashed due to the production of cells with negative densities. The time per step was estimated from averaging over $0.1\,T$ within the
last turnover time, i.e., between $3.2\leq t/T\leq3.3$ by averaging over 309, 70 and 71 steps for the \textsf{FLASH-Roe}, \textsf{HLL3R} and \textsf{HLL5R}, respectively to avoid variations produced by I/O processing. Considering the time per step, \textsf{HLL3R} and \textsf{HLL5R} are about 4\% and 3\%, respectively faster than \textsf{FLASH-Roe}. This speed-up of \textsf{HLL3R} is slightly less than the speed-up measured for the Orszag-Tang test (cf.~Tab.~\ref{tab:clock}). The difference between \textsf{HLL3R} and \textsf{HLL5R} is almost negligible. Indeed, the two corresponding solvers have a very similar structure, except that the HLL5R solver typically goes through one more if-test compared to the HLL3R solver. HLL5R goes through a few more time steps, which is either because of its slightly more restrictive CFL constraint, or flow features such as local differences Alfv{\'e}n speed.

\begin{table}[tbp]
\begin{center}
\def\arraystretch{1.1}
\begin{tabular}{lrrrr}
\hline
\hline
Scheme & Time per step $\left[\s\right]$ & Total steps & Total CPU hours & CFL\\
\hline
\textsf{FLASH-Roe}    &  $69.8\pm0.5$  &  16,502  &  20,480  &  0.2 \\
\textsf{HLL3R}  &  $67.3\pm0.5$  &   1,920  &   2,297  &  0.8 \\
\textsf{HLL5R}  &  $67.9\pm0.5$  &   2,065  &   2,493  &  0.8 \\
\hline
\\
\end{tabular}
\caption{Performance of the MHD schemes in driven MHD turbulence with $\Ma=2$ for a numerical resolution of $256^3$ grid cells. All runs were performed in a mode of parallel computation with 64 CPUs on the SGI Altix 4700 (HLRB II) at the Leibniz Supercomputing Center in Munich. Note that the run with the \textsf{FLASH-Roe} scheme was unstable for CFL numbers $>0.2$ in this test.}
\label{tab:turb_cputime}
\end{center}
\end{table}

The total number of steps and the total CPU time spent by $t_\mathrm{end}=3.3\,T$ for each run are shown in the third and fourth column of Table~\ref{tab:turb_cputime}. Given the factor of 4 between the CFL number used with \textsf{FLASH-Roe} compared to \textsf{HLL3R} and \textsf{HLL5R} it is not surprising that the total number of steps is much higher for \textsf{FLASH-Roe}. However, the number of steps for \textsf{FLASH-Roe} is about 8.6 and 8.0 times higher compared to \textsf{HLL3R} and \textsf{HLL5R} respectively, which is more than the difference in the CFL numbers. The reason for this is that \textsf{FLASH-Roe} is close to being unstable even with CFL = 0.2. This produces timesteps that are often significantly smaller than necessary. Taken altogether, for the problem setup discussed in this section, \textsf{HLL3R} and \textsf{HLL5R} are factors of 8.9 and 8.2 respectively faster than \textsf{FLASH-Roe}, which is reflected by the total amount of CPU time spent for a successful MHD turbulence run with $\Ma\approx2$. For higher Mach number runs, however, \textsf{FLASH-Roe} was extremely unstable, which did not allow us to make this comparison of the schemes at higher Mach number.

\subsection{Resolution study} \label{sec:turb_resol}
As outlined in the introduction, interstellar turbulence is highly supersonic with typical Mach numbers of the order of $\Ma\approx10$. As discussed in the comparison of the different MHD schemes in the previous subsections, the \textsf{FLASH-Roe} scheme turned out to be highly unstable even for low Mach number turbulence. This limitation is overcome with the \textsf{HLL3R} and \textsf{HLL5R} schemes. In this section, we apply our new \textsf{HLL3R} scheme to MHD turbulence with $\Ma=10$ as an example of the high Mach number turbulence typical of the interstellar medium and discuss numerical resolution issues. The three runs analysed in this section were run at numerical resolutions of $128^3$, $256^3$ and $512^3$ grid cells with a slightly smaller initial magnetic field, $B_0=4.4\times10^{-6}\,\G$, which gives an initial plasma beta of $\beta\approx1$.

\begin{figure}[tbp]
\centering
\includegraphics[width=0.45\linewidth]{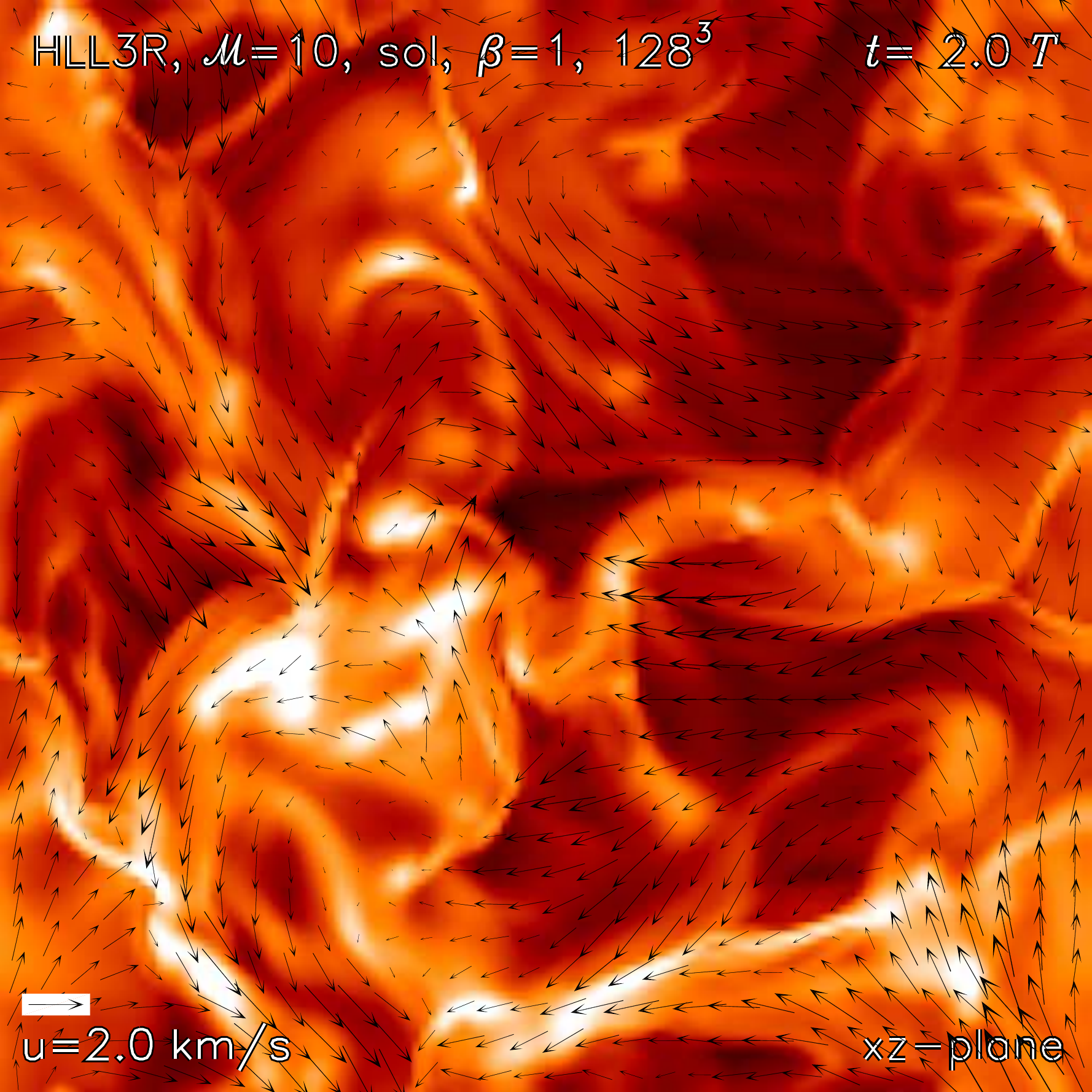}
\includegraphics[width=0.45\linewidth]{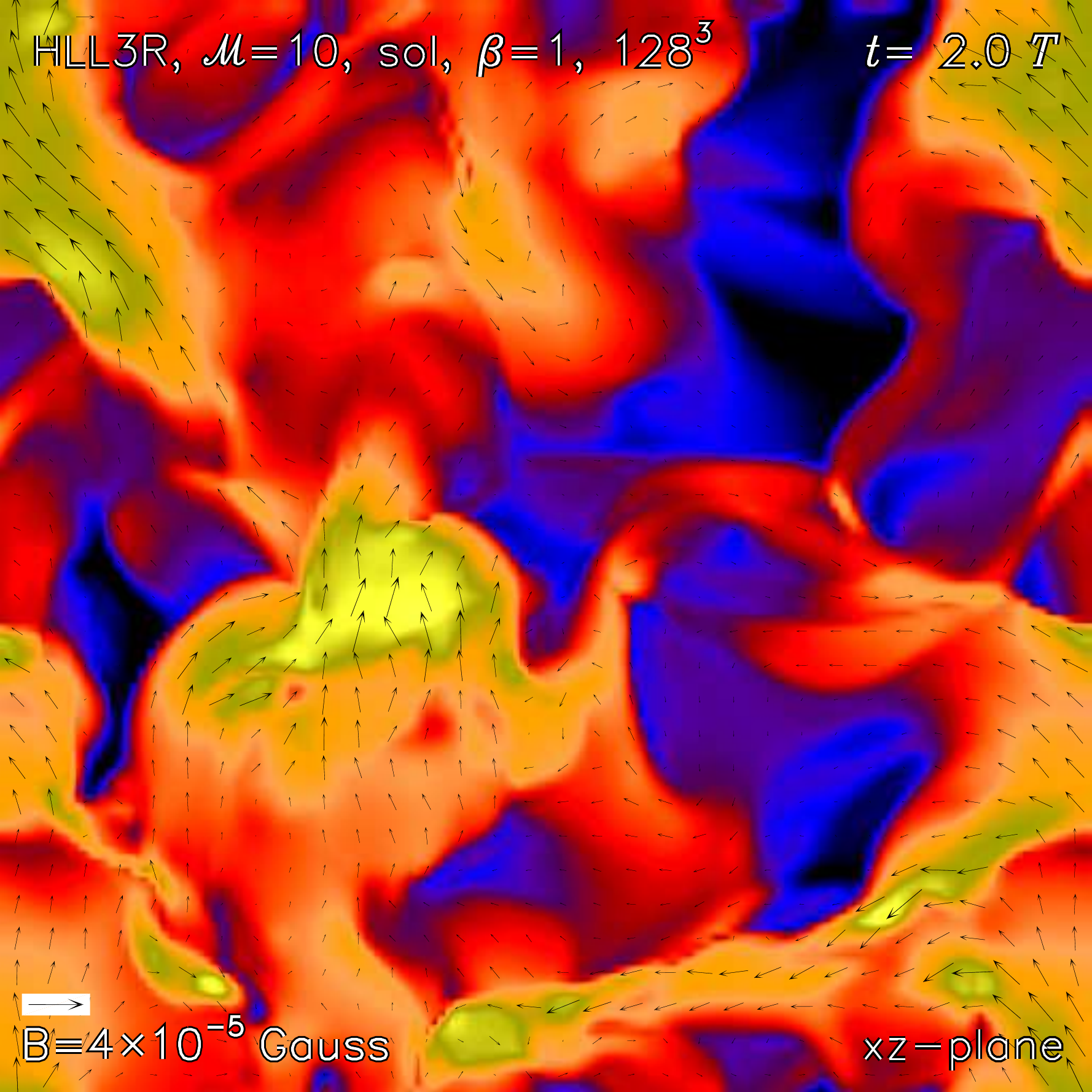}
\includegraphics[width=0.45\linewidth]{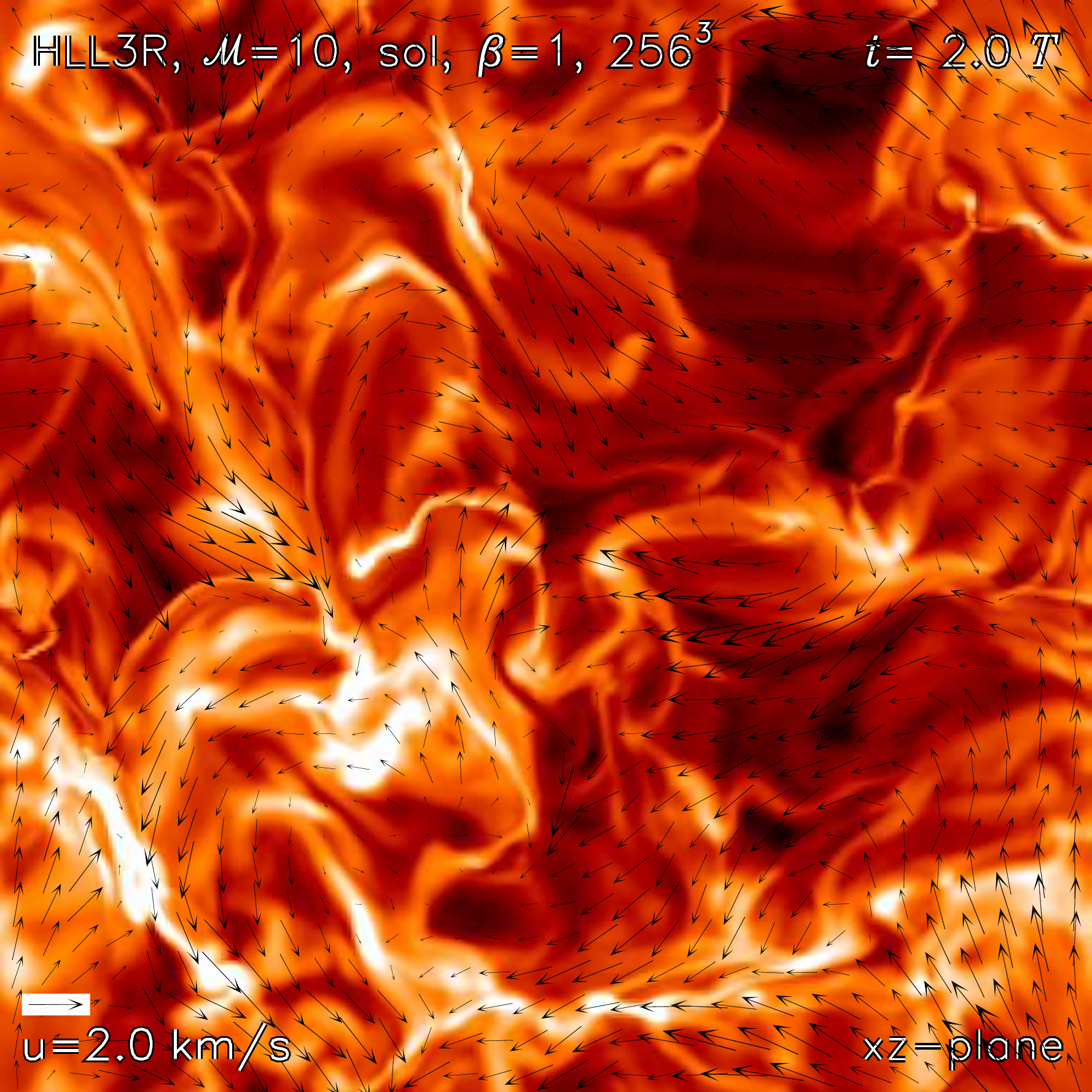}
\includegraphics[width=0.45\linewidth]{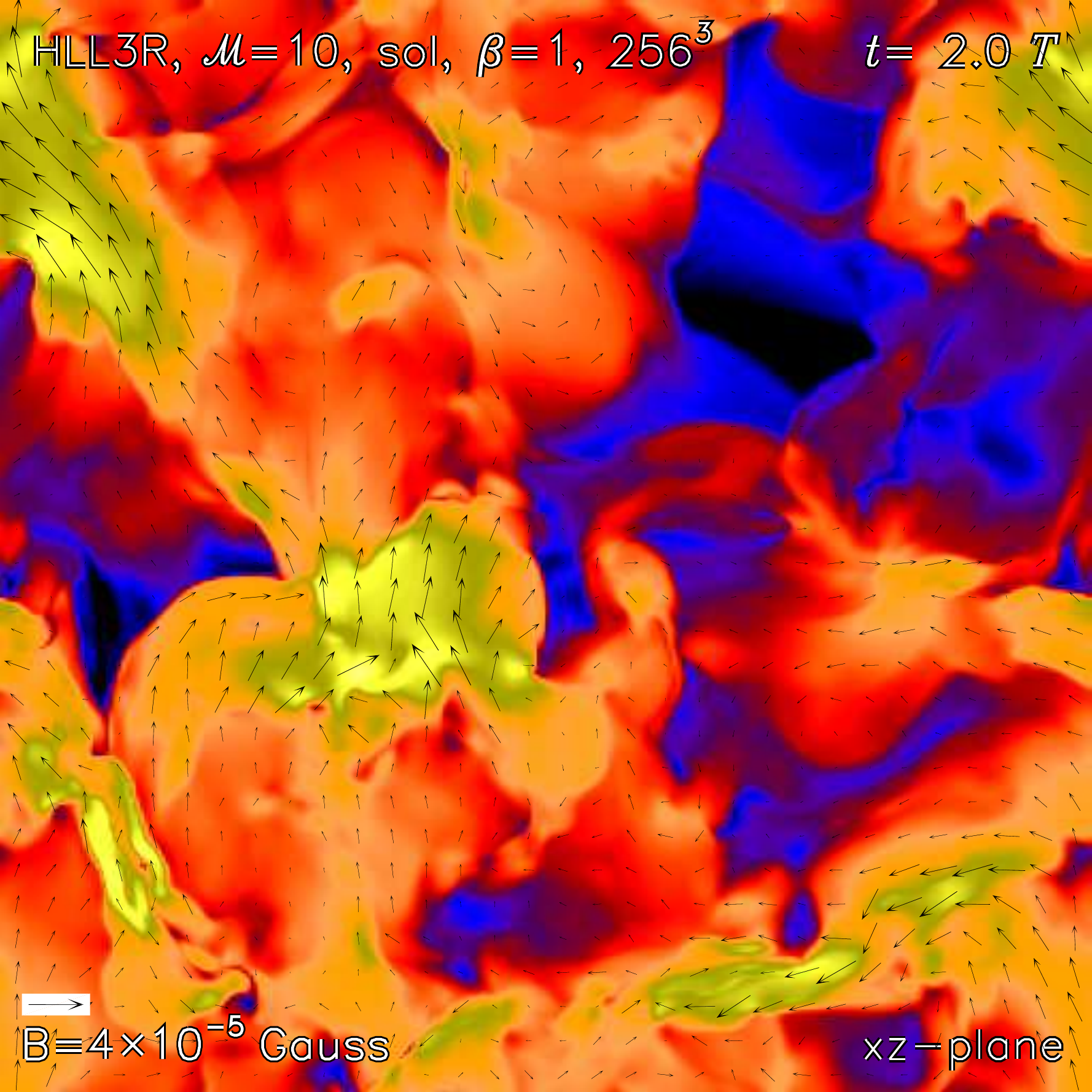}
\includegraphics[width=0.45\linewidth]{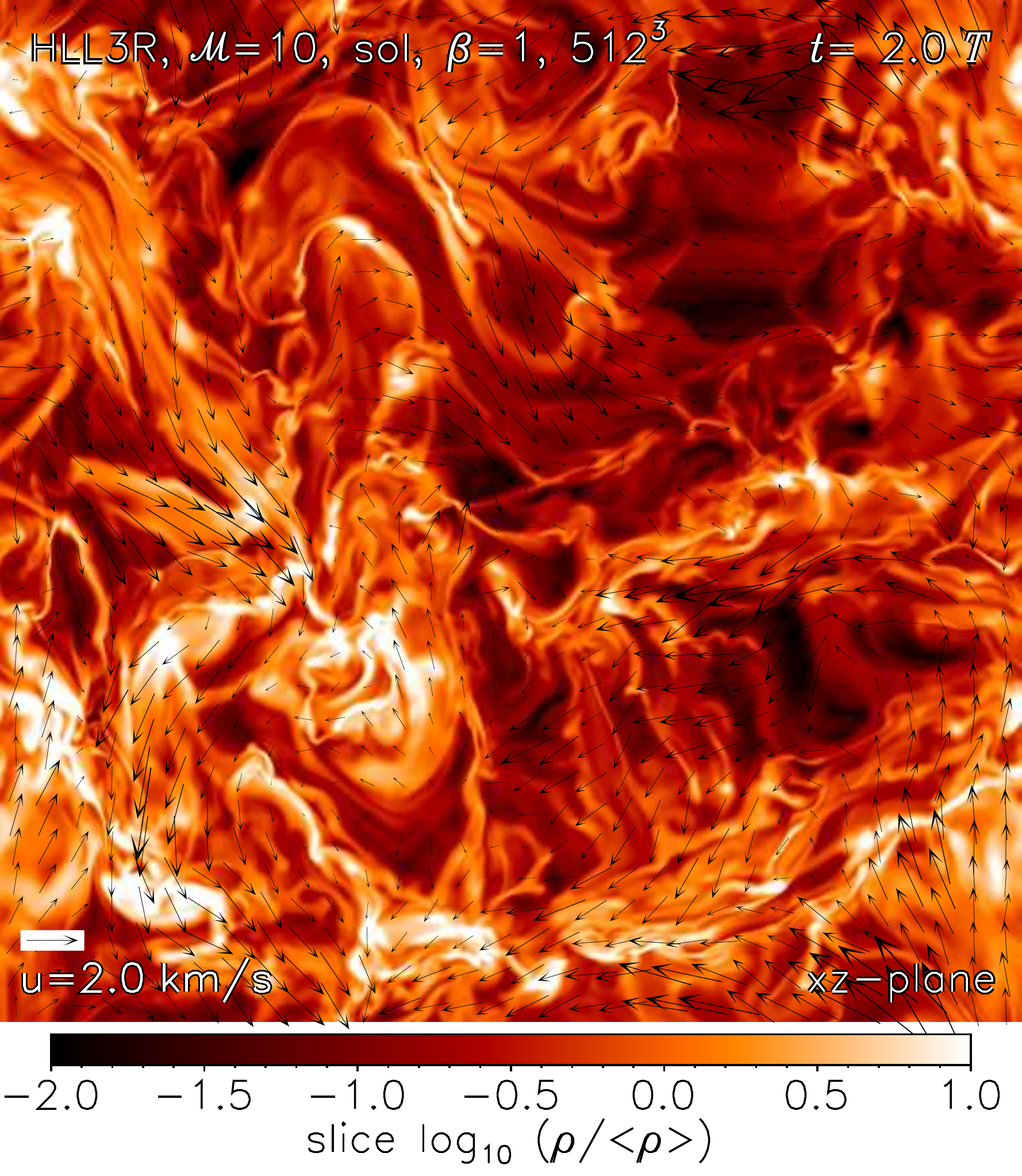}
\includegraphics[width=0.45\linewidth]{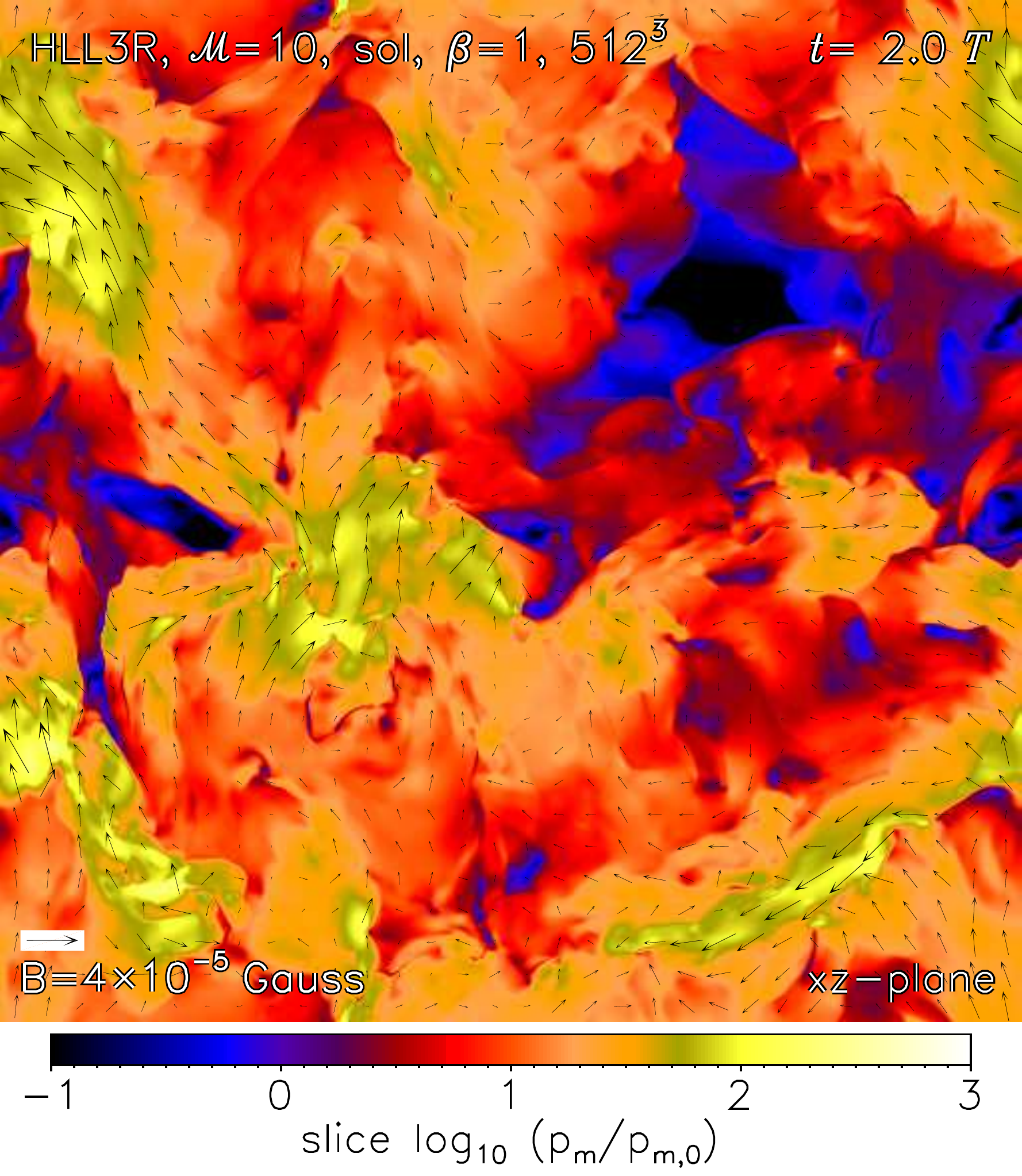}
\caption{Slices of the density (\emph{left}) and magnetic pressure (\emph{right}) at $y=L/2$ and $t=2\,T$ for numerical resolutions of $128^3$ (\emph{top}), $256^3$ (\emph{middle}), and $512^3$ (\emph{bottom}).}
\label{fig:turb_snapshots}
\end{figure}

First, we show slices through the three-dimensional box at $y=L/2$ of the density and magnetic pressure fields in Figure~\ref{fig:turb_snapshots}. With increasing resolution more small-scale structure is resolved. The density and magnetic pressure are correlated due to the compression of magnetic field lines in shocks. Vector fields showing the velocity and the magnetic field structure are overlaid on the slices of the density and magnetic pressure, respectively.

\begin{figure}[tbp]
\centering
\includegraphics[width=0.6\linewidth]{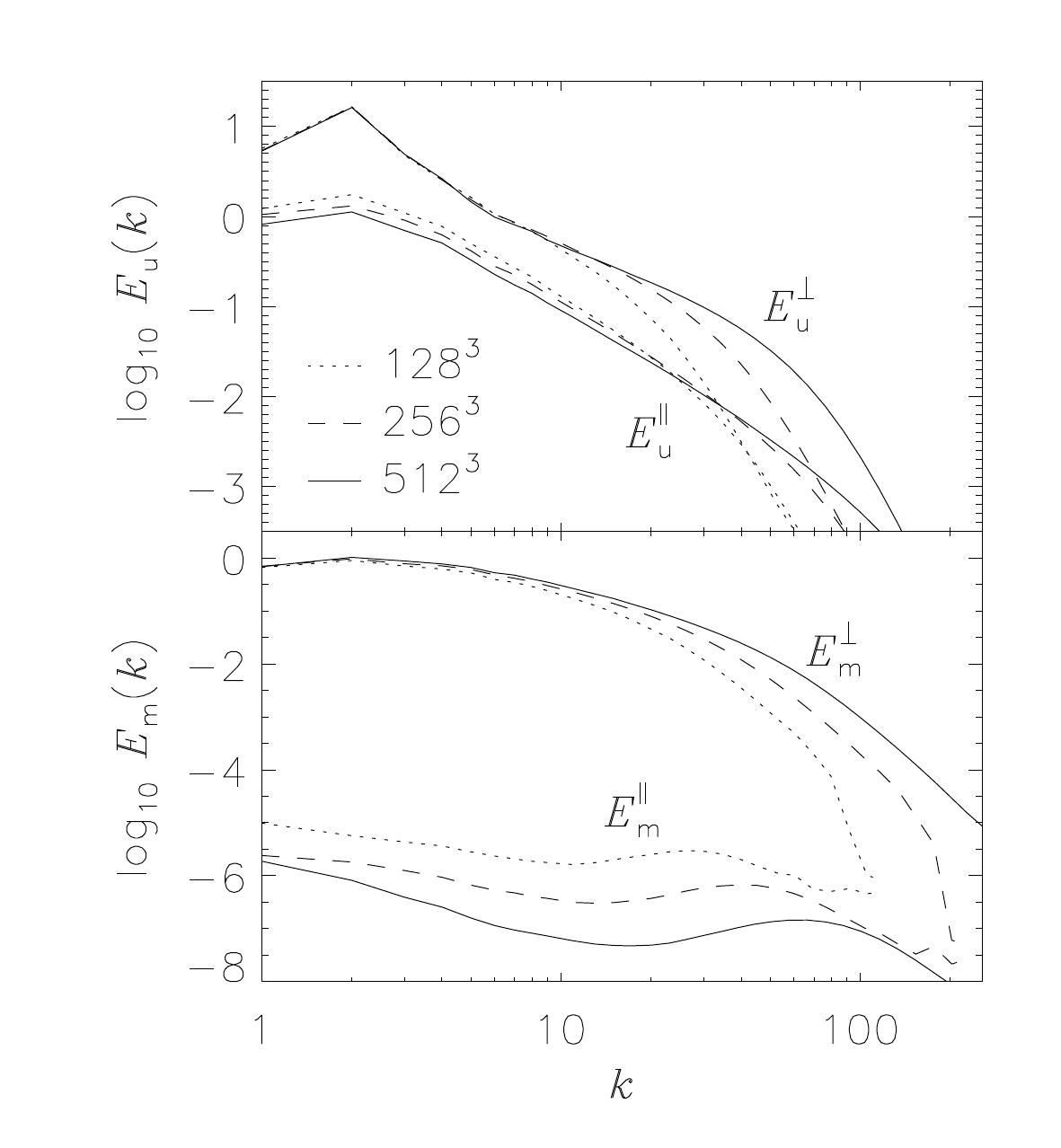}
\caption{Comparison of the Fourier spectra of the turbulent velocity (\emph{top panel}) and the turbulent magnetic field (\emph{bottom panel}) for \textsf{HLL3R} in $\Ma=10$, $\beta=1$ MHD turbulence. The influence of the grid resolution is shown: $128^3$ (dotted), $256^3$ (dashed) and $512^3$ (solid). Both the velocity and the magnetic field spectra were decomposed into their rotational and compressible parts, $E^\perp$ and $E^\parallel$, respectively, as in Figure~\ref{fig:turb_spect}.}
\label{fig:turb_spect_resol}
\end{figure}

Numerical dissipation becomes smaller with higher resolution. Since the inertial range of the turbulence is defined as all the length scales much smaller than the forcing scale and much larger
than the dissipation scale, a minimum numerical resolution of about $512^3$
grid cells is required to separate the inertial range from the forcing
and dissipative scales. To see how dissipation depends on numerical resolution we plot the velocity and magnetic field spectra in Figure~\ref{fig:turb_spect_resol} for the three different resolutions in analogy to Figure~\ref{fig:turb_spect}. The turbulence is driven on large scales corresponding to $k\approx2$ (injection scale, $L_\mathrm{inj}$), while the spectra become steeper and begin dropping to zero at wavenumbers $k\gtrsim10$, $20$, and $40$ for $128^3$, $256^3$, and $512^3$, indicating the onset of dissipation (on scales $\ell_\mathrm{dis}$), respectively. Thus, the range of scales separating the injection scale from the dissipation scale increases with increasing resolution. However, the inertial range is defined as all scales, $\ell$ with $L_\mathrm{inj}\gg\ell\gg\ell_\mathrm{dis}$. Purely hydrodynamical simulations of driven turbulence with up to $1024^3$ grid cells show that the inertial range just becomes apparent for resolutions $\gtrsim512^3$, which means that we cannot see a convergence of the slope in the spectra shown in Figure~\ref{fig:turb_spect_resol}. However, measuring the power-law slope, $E\propto k^\beta$ of the velocity spectra in the wavenumber range $5\lesssim k\lesssim10$ at $512^3$ yields slopes of $\beta\approx-1.6\pm0.1$, and $\beta\approx-1.9\pm0.1$, for $E_\mathrm{u}^\perp$ and $E_\mathrm{u}^\parallel$, respectively. The solenoidal part, $E_\mathrm{u}^\perp$ is in good agreement with the Kolmogorov spectrum ($\beta=-5/3$) \cite{Kolmogorov1941c,Frisch1995}, while the compressible part, $E_\mathrm{u}^\parallel$ is significantly steeper and closer to Burgers turbulence ($\beta=2$) \cite{Burgers1948}, applicable to a shock-dominated medium. Note that $E_\mathrm{m}^\parallel$ clearly decreases with increasing resolution.

\subsection{Turbulent dynamo}
\label{dynamo}
Magnetic fields are ubiquitous in molecular clouds, but it remains controversial whether these fields have an influence on the cloud dynamics (see, e.g., \cite{MacLowKlessen2004}). However, it is widely accepted that magnetic fields play a significant role on the scales of protostellar cores, where they lead to the generation of spectacular jets and outflows, launched from the protostellar disks, a process for which a wound-up magnetic field seems to be the key \cite{BlandfordPayne1982,LyndenBell2003}. Thus, we test in this section whether our new MHD schemes can represent wound-up magnetic field configurations with the same turbulence-in-a-box approach as in the previous subsections.

The magnetic pressure can become comparable to the thermal pressure in dense cores due to the amplification of the magnetic field through first, compression of magnetic field lines, and second, due to the winding, twisting and folding of the field lines by vorticity, a process called turbulent dynamo (see \cite{BrandenburgSubramanian2005} for a comprehensive review of turbulent dynamo action in astrophysical systems). Magnetic field amplification in the early universe during the formation of the first stars and galaxies was discussed analytically in \cite{SchleicherEtAl2010}, showing that the magnetic pressure can reach levels of about 20\% of the thermal pressure in primordial mini-halos, thus potentially influencing the fragmentation of the gas. Numerical simulations also show that the turbulent dynamo is an efficient process to amplify small seed magnetic fields during the formation of the first stars and galaxies (\cite{Xu2009,Sur2010}). In present-day star formation the influence of magnetic fields on the scales of dense cores were investigated in numerical work (\cite{PriceBate2007,HennebelleTeyssier2008}), concluding that magnetic fields strongly affect fragmentation of dense gas. Understanding the magnetic field growth due to the turbulent dynamo is thus crucial for future studies of star formation. Many studies have focused on subsonic turbulence (e.g., \cite{SchekochihinEtAl2004}), with only very few contributions on the supersonic regime. For instance, \cite{HaugenBrandenburgMee2004} only studied the turbulent dynamo for mildly supersonic Mach numbers, $\Ma\lesssim2.5$. For molecular clouds however, the highly supersonic regime is more relevant.

\begin{figure}[tbp]
\centering
\includegraphics[width=0.6\linewidth]{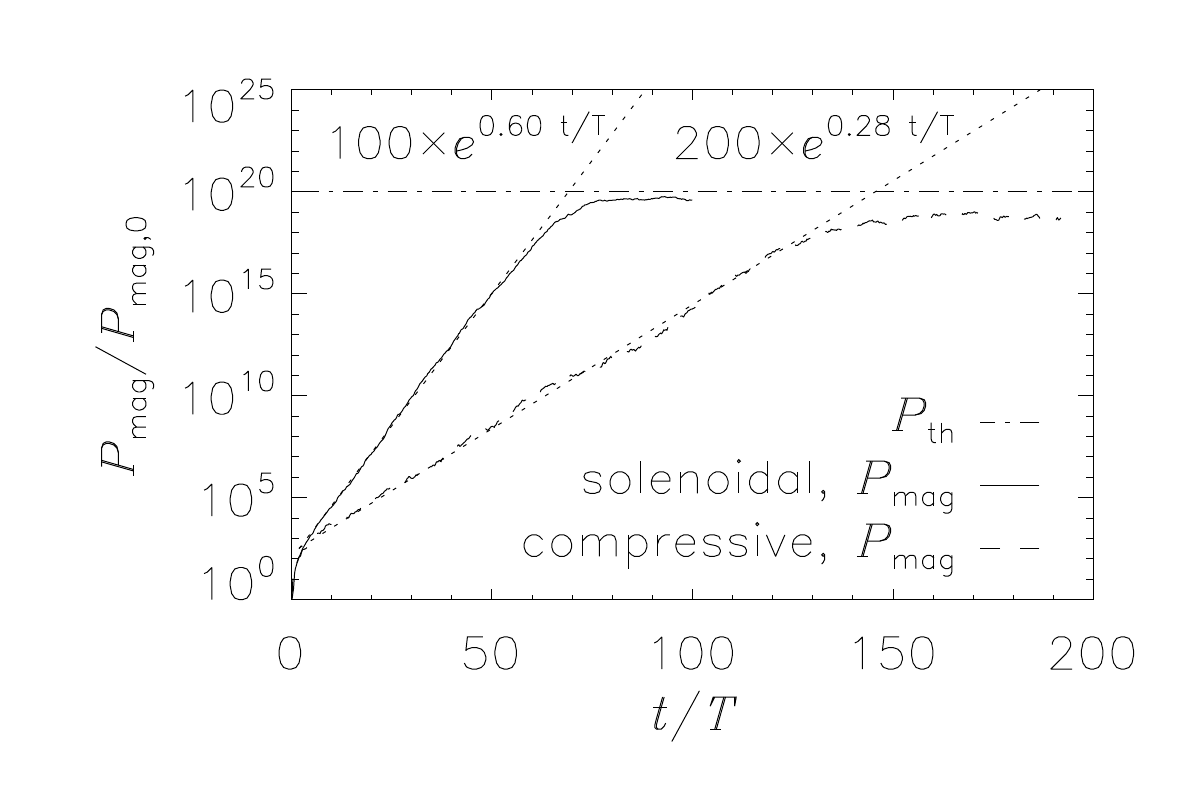}
\caption{Shows the magnetic pressure as a function of time (for 200 eddy turnover times, $T$) for solenoidal forcing (solid line) and compressive forcing (dashed line). The dash-dotted horizontal line shows the thermal pressure. The turbulent dynamo works with an exponential growth rate of about $0.60/T$ for solenoidal forcing and $0.28/T$ for compressive forcing (dotted lines). For comparison, typical amplification rates found in subsonic, solenoidally driven turbulence are about $0.5/T$ (e.g., \cite{SchekochihinEtAl2004}).}
\label{fig:turb_dynamo}
\end{figure}

Figure~\ref{fig:turb_dynamo} shows a comparison study of the turbulent dynamo operating in the supersonic regime ($\Ma=5$). We used the \textsf{HLL3R} for this test with a numerical resolution of $128^3$ grid cells. Unlike the previous turbulence runs discussed above we started from an extremely small initial magnetic field, $B_0=4.4\times10^{-16}\,\G$, which corresponds to an initial plasma beta of $\beta\approx10^{20}$. We used two limiting cases to drive the turbulence: purely solenoidal (divergence-free) forcing as above, and purely compressive (curl-free) forcing (see \cite{FederrathDuvalKlessenSchmidtMacLow2010} for a detailed analysis of the differences of solenoidal and compressive turbulence forcings and their role in the context of molecular cloud dynamics and star formation). Figure~\ref{fig:turb_dynamo} shows that the turbulent dynamo leads to an exponential growth of the magnetic pressure over more than 15 orders of magnitude within about 100 large-scale eddy turnover times, $T$. The dynamo growth rate is about twice as large for solenoidal forcing ($\approx0.60/T$) compared to compressive forcing ($\approx0.28/T$), due to the higher average vorticity generated by solenoidal forcing (see \cite{FederrathDuvalKlessenSchmidtMacLow2010}), which makes the small-scale dynamo more efficient. The actual growth rates, however, depend on the kinematic and magnetic Reynolds numbers. Since we did not add physical viscosity and magnetic diffusivity, these numbers are controlled by numerical viscosity and diffusivity, and thus by numerical resolution. A resolution study of the dynamo in recent FLASH simulations is presented in \cite{Sur2010} and in \cite{Federrath2011}, the kinematic and magnetic Reynolds numbers found in dynamo simulations similar to the ones presented here are around 200 for a numerical resolution of $128^3$ grid cells. A more detailed analysis of the Mach number dependence of the turbulent dynamo amplification in solenoidal and compressive forcings is in preparation. The dynamo saturates after about $70\,T$ and $140\,T$ for solenoidal and compressive forcing, respectively. The saturation level is close to the thermal pressure in both cases. For compressive forcing the saturated magnetic pressure is about 5\% of the thermal pressure, while it is 40\% for solenoidal forcing. This numerical test shows that the turbulent dynamo works with the new \textsf{HLL3R} scheme. This is an important test as it shows that the scheme reproduces the expected amplification of the magnetic pressure due to the winding-up of magnetic field lines.


\section{Summary}
We presented an implementation of an accurate, efficient and highly stable numerical method for MHD problems. The method is reviewed here, and presented in detail in \cite{W1}. It is implemented as a modification of the FLASH code \cite{FLASH}, which enables large-scale, multi-processor simulations and adaptive mesh refinement. In \cite{W1} it was found that our method could handle significantly larger ranges of the sonic Mach number and plasma $\beta$ than a standard MHD scheme. This was confirmed in this paper by comparisons with the standard FLASH code. The algorithmic changes underlying the increased stability can be broken down into three parts:
\begin{enumerate}
\item
An entropy stable approximate Riemann solver that preserves positivity of density and internal energy (\cite{BKW1,BKW2}).
\item
For second order accuracy, a reconstruction method that ensures positivity (\cite{W1}).
\item
In multidimensions, a stable discretisation of the Powell system  (\cite{W1}).
\end{enumerate}
All these ingredients were essential in obtaining the desired stability and efficiency of the overall scheme. The different elements of the new scheme have been studied separately in previous papers. While \cite{W1} focused on the positive second-order algorithm and multidimensionality, only a single approximate Riemann solver,  HLL3R, was considered. The present study contrasts a standard scheme to the combination of these three new ingredients.

The new scheme was implemented in two versions featuring the 3-wave (HLL3R) and 5-wave (HLL5R) approximate Riemann solvers of \cite{BKW2} respectively, while there were two standard implementations of the FLASH code (version 2.5), using the Roe and the HLLE approximate Riemann solvers.  We observed some increase of numerical dissipation compared to the Roe solver of FLASH, but it was minor, and due to the replacement of the Roe solver with the robust and efficient HLL-type solvers. The HLLE solver was found to be the most dissipative, while HLL5R showed almost identical dissipation properties and accuracy to the Roe solver. HLL3R was ranked between HLLE and Roe in terms of accuracy. 

As a physical application, we have considered forced MHD turbulence at high sonic Mach number. We were able to compare the new and old schemes at RMS sonic Mach number 2 with an initial plasma $\beta=0.25$. The schemes were all found to give similar and reasonable results, but the new schemes HLL3R and HLL5R were altogether about eight times more efficient in this test. The Roe solver-based scheme in the FLASH code was slightly less dissipative, but had to be run at a four times lower CFL number to be stable. At RMS sonic Mach number 10 only the new schemes yielded physical results. We found reasonable dependence of dissipation on numerical resolution at this Mach number, and were able to infer a small inertial range from the velocity power spectra at $512^3$ resolution. Finally, we studied the turbulent dynamo action at RMS Mach number 5 with the new scheme. So far, there have been very few studies of turbulent dynamos in the supersonic regime. We found dynamo-generated exponential growth rates of the magnetic pressure that differed according to the type of forcing mechanism, i.e., solenoidal versus compressive forcing \cite{FederrathKlessenSchmidt2008,FederrathKlessenSchmidt2009,FederrathDuvalKlessenSchmidtMacLow2010}. Turbulence simulations with the new scheme across a wide range of modest to large sonic and Alfv{\'e}nic Mach numbers have been presented in \cite{BruntFederrathPrice2010a,BruntFederrathPrice2010b}.

The two relaxation-based approximate Riemann solvers HLL3R and HLL5R have previously not been compared at high order and in higher spatial dimensions than 1D. In many cases we found them to give similar results, with HLL3R being slightly more efficient. However, we presented one case, a 2D Kelvin-Helmholtz instability, where the more detailed HLL5R was significantly less viscous. This was because of velocity shears parallel to both the grid and the magnetic field lines. In a three-dimensional turbulence run at Mach 2, we also found HLL5R to be somewhat less dissipative than HLL3R, giving results that were very close to the Roe solver-based FLASH version. 

The new FLASH MHD module is freely available upon contact with the corresponding author.


\end{document}